\documentclass[12pt, a4paper]{article}
\usepackage{amsmath}
\usepackage{multirow}
\usepackage{amsfonts}
\usepackage{amssymb,graphics,psfrag}
\usepackage{array,epsfig,multirow,graphicx}
\usepackage{comment}
\usepackage{slashed}
\usepackage{hyperref}
\usepackage{tensor}
\usepackage{booktabs}
\usepackage{colortbl}
\usepackage{hhline}
\usepackage{mathtools}
\usepackage{enumitem}
\usepackage{mathrsfs}  
\usepackage{xfrac}
\usepackage{tikz}
\usepackage{amssymb}
\usepackage[numbers,sort&compress]{natbib}
\usetikzlibrary{decorations.markings}
\usetikzlibrary{shapes,arrows}
\usetikzlibrary{decorations.pathreplacing}
\usetikzlibrary{arrows,positioning}
\usepackage{verbatim}
\usepackage{makecell}
\usepackage{setspace}
\usepackage{anyfontsize}
\usepackage{ctable}
\usepackage{braket}
\usepackage{arydshln}
\usepackage{geometry}

\numberwithin{equation}{section}
\numberwithin{table}{section}

\newcommand{\beq}{\begin{equation}}
\newcommand{\eeq}{\end{equation}}
\newcommand{\be}{\begin{equation}}
\newcommand{\ee}{\end{equation}}
\newcommand{\bea}{\begin{eqnarray}}
\newcommand{\eea}{\end{eqnarray}}
\newcommand{\ben}{\begin{eqnarray*}}
\newcommand{\een}{\end{eqnarray*}}               
\newcommand{\ba}{\begin{align}}
\newcommand{\ea}{\end{align}}
\newcommand{\bt}{\begin{tabular}}
\newcommand{\et}{\end{tabular}}
\newcommand{\bc}{\begin{center}}
\newcommand{\ec}{\end{center}}
\newcommand{\ax}{\alpha}
\newcommand{\bx}{\beta}

\newcommand{\cO}{\mathcal{O}}

\newcommand{\cN}{\mathcal{N}}

\newcommand{\cA}{\mathcal{A}}

\newcommand{\cR}{\mathcal{R}}

\newcommand{\nn}{\nonumber}

\newcommand{\cref}{{\bf [check ref]}}

\newcommand{\tr}{\mathrm{tr}}

\definecolor{mppgreen}{RGB}{17,102,86}
\definecolor{mppgray}{RGB}{221,222,214}

\def\tbz{{\scriptscriptstyle{(0)}}}

\def\fr{\frac}

\def\Z0{Z^\tbz}

\def\t{\text}

\def\w{\wedge}
\def\tfr{\tfrac}

\def\tr{\text{tr}\,}
\def\cR{\mathcal{R}}

\def\w{\wedge}
\def\cR{\mathcal{R}}

\def\lab{\label}

\newcommand{\RR}{\boldsymbol{\mathrm{R}}}

\newcommand{\dd}{\mathrm{d}}

\def\blfootnote{\xdef\@thefnmark{}\@footnotetext}
\long\def\symbolfootnote[#1]#2{\begingroup%
\def\thefootnote{\fnsymbol{footnote}}\footnote[#1]{#2}\endgroup}

\newcolumntype{L}[1]{>{\raggedright\let\newline\\\arraybackslash\hspace{0pt}}m{#1}}
\newcolumntype{C}[1]{>{\centering\let\newline\\\arraybackslash\hspace{0pt}}m{#1}}
\newcolumntype{R}[1]{>{\raggedleft\let\newline\\\arraybackslash\hspace{0pt}}m{#1}}
\renewcommand{\arraystretch}{1.5}

\makeatletter
\DeclareRobustCommand*{\bfseries}{%
   \not@math@alphabet\bfseries\mathbf
   \fontseries\bfdefault\selectfont
   \boldmath
}
\makeatother

\usepackage[labelfont=bf]{caption}

\begin{document}

\baselineskip=15pt

\begin{titlepage}
\begin{flushright}
\parbox[t]{1.8in}{\begin{flushright} ~ \\
~ \end{flushright}}
\end{flushright}

\begin{center}

\vspace*{ 1.2cm}

\begin{spacing}{2.5}
\bf{{{\fontsize{21}{1}\selectfont  Anomalies of (0,4) SCFTs from F-theory}}}
\end{spacing}

\vskip 1.2cm

\renewcommand{\thefootnote}{}
\begin{center}
 {Christopher Couzens, Huibert het Lam, Kilian Mayer and Stefan Vandoren 

 \footnotetext{c.a.couzens@uu.nl~ \ ~ h.hetlam@uu.nl~ \ ~ k.mayer@uu.nl~ \ ~  s.j.g.vandoren@uu.nl}}
\end{center}
\vskip .2cm

{
Institute for Theoretical Physics and \\
Center for Extreme Matter and Emergent Phenomena,\\
Utrecht University, Princetonplein 5, 3584 CE Utrecht, The Netherlands\vspace{0.5cm}
}

\vspace*{.2cm}

\end{center}

 \renewcommand{\thefootnote}{\arabic{footnote}}
 
\begin{center} {\bf Abstract } \end{center}

We study the macroscopics of 2d $\mathcal{N}=(0,4)$ SCFTs arising from F-theory constructions. The class of 2d SCFTs we consider live on black strings which are obtained by wrapping D3-branes on a curve in the base of a possibly singular elliptically fibered Calabi--Yau threefold. In addition, we allow the D3-branes to probe ALE or ALF spaces transversely.  We compute anomaly coefficients of these SCFTs by determining Chern--Simons terms in the 3d action resulting from the reduction of 6d $\mathcal{N}=(1,0)$ supergravity on the compact space surrounding the black string. Essential contributions to these coefficients are from one-loop induced Chern--Simons terms arising from integrating out massive Kaluza--Klein modes.

\end{titlepage}

\tableofcontents

\newpage


\section{Introduction}

In string theory an oft-used approach to studying black holes is to consider wrapped black strings in one dimension higher. The supergravity solution of the black string contains an AdS$_3$ factor in the near-horizon limit, and via the AdS/CFT corespondence there exists a dual two-dimensional CFT. This CFT gives a microscopic interpretation of the entropy of the black string and by reduction also that of the black hole. The black string arises from a brane configuration wrapped in the internal geometry of an M/F- or string theory compactification and the CFT is the infrared worldvolume theory living on these wrapped branes. The prime example of studying black holes in this way is given by the two-dimensional $\mathcal{N}=(4,4)$ SCFT living on the D1-D5 system in type IIB string theory on $\text{T}^4\times \text{S}^1$ \cite{Strominger:1996sh}. A further example is provided by the MSW string which lives in M-theory on a Calabi--Yau threefold (CY$_3$), and is obtained by wrapping M5-branes on a divisor in the CY$_3$. In the infrared the worldvolume theory living on the resulting string flows to a 2d $\mathcal{N}=(0,4)$ SCFT.  Though little is known about this SCFT, its central charge was computed in \cite{Maldacena:1997de} and gives a microscopic interpretation of the black hole entropy via Cardy's formula. There are closely related constructions in F-theory, where D3-branes wrap a curve in the base of an elliptically fibered CY$_3$, again leading to $\mathcal{N}=(0,4)$ SCFTs in the infrared \cite{Vafa:1997gr}. In this example the black strings live in the six-dimensional $\mathcal{N}=(1,0)$ supergravity theory obtained by reducing F-theory on the CY$_3$ and have an AdS$_3\times$S$^3$ near-horizon. These SCFTs not only have an $SU(2)_R$ symmetry in the supersymmetric right-moving sector, but also an $SU(2)_L$ symmetry in the left-moving sector. The level of this $SU(2)_L$ current algebra was computed in \cite{Haghighat:2015ega} and plays a prominent role in Cardy's formula which determines the entropy of the five-dimensional spinning black holes obtained by wrapping these black strings on a circle.

The study of supersymmetric solutions admitting AdS$_3$ factors, not necessarily related to black strings but more generally for AdS/CFT purposes, has a long and rich history. Various works classify and identify such solutions. Of interest to us in this paper are those which preserve $\mathcal{N}=(0,4)$ supersymmetry. In M-theory, this program was initiated in \cite{Martelli:2003ki} and later refined in \cite{Colgain:2010wb}. More recently additional solutions have been found in \cite{Lozano:2020bxo} where earlier works on $\mathcal{N}=(0,4)$ AdS$_3$ solutions in (massive) type IIA \cite{Lozano:2019emq,Lozano:2019jza,Lozano:2019zvg,Lozano:2019ywa} were used. Further work in type IIA (and also type IIB) can be found in \cite{Macpherson:2018mif} where solutions preserving the large $\mathcal{N}=(0,4)$ supersymmetry algebra were found.\footnote{See also \cite{Dibitetto:2018ftj} for solutions with exceptional supersymmetry algebras.} Type IIB and F-theory $\mathcal{N}=(0,4)$ solutions were discussed in \cite{Couzens:2017way}, whilst further work on AdS$_3$ solutions in type IIB, with varying amounts of supersymmetry, can be found in \cite{Kim:2005ez,Gauntlett:2006ns, Gauntlett:2006af,Donos:2008ug,Eberhardt:2017uup,Couzens:2017nnr,Passias:2019rga,Couzens:2019iog,Couzens:2019mkh, Legramandi:2019xqd}. 

Typical observables which are computed are the central charges and levels of current algebras of the CFT. They can provide evidence for the AdS/CFT correspondence and determine the entropy when the setup descends from a black string. Various methods and directions have been used to compute these observables. Knowledge of the UV data of the CFT is often sufficient by virtue of 't Hooft anomaly matching. Anomalies of the UV theory can be determined directly using the spectrum, by reducing the anomaly polynomial of a higher-dimensional parent theory or by anomaly inflow \cite{Shimizu:2016lbw,Kim:2016foj,Kim:2019vuc,DelZotto:2018tcj,DelZotto:2016pvm}. The central charges can then be computed either via spectrum counting or using c-extremization \cite{Benini:2012cz}\footnote{Strictly in order to perform c-extremization one needs only $\mathcal{N}=(0,2)$ supersymmetry and c-extremization is often not necessary when the R-symmetry is a non-abelian group since generically it cannot mix with other symmetries. However, there are cases where a non-abelian R-symmetry can mix with other symmetries and application of c-extremization using the $U(1)_{R}$ of the Cartan of the symmetry group is necessary, see for example \cite{Tong:2014yna}.}. Alternatively, the central charges and levels can be determined using holographic methods in the dual supergravity theory. 

Throughout this paper we are interested in $\mathcal{N}=(0,4)$ SCFTs dual to near-horizon geometries of the form AdS$_3 \times \mathrm{S}^3/\Gamma$, where $\Gamma$ is a freely acting finite subgroup of $SU(2)_L$. Such a setup in F-theory was discussed in \cite{Bena:2006qm,Grimm:2018weo}. There D3-branes probe a transverse Taub-NUT space and are wrapped on both a curve in the base of a smooth elliptically fibered Calabi--Yau threefold CY$_3$ and a circle. The transverse Taub-NUT space leads to a $\Gamma=\mathbb{Z}_m$ quotient of the three-sphere in the near-horizon. Since Taub-NUT space asymptotically approaches $\mathbb{R}^3\times \text{S}^1$ there are two circles in the geometry on which one can compactify, allowing for the construction of four-dimensional black holes in F-theory. Two extensions are possible here and are the subject of this paper. The first is to consider non-smooth Calabi--Yau threefolds. This allows for additional (non-abelian) gauge groups in the supergravity setup, which act as flavour symmetries in the dual $\mathcal{N}=(0,4)$ SCFT. We macroscopically determine the levels of the corresponding current algebras in the SCFT. The second extension we consider is to allow for more general subgroups $\Gamma$. In fact one can generalize even further by allowing the transverse space probed by the D3-brane to be either asymptotically locally flat (ALF) (to which Taub-NUT belongs) or asymptotically locally Euclidean (ALE). Both families of spaces allow for an AD(E) classification which in the near-horizon leads to a $\Gamma$ quotiented three-sphere. This extension was considered in type IIB on K3 in \cite{Couzens:2019wls} (see also earlier work in \cite{Banerjee:2009uk,Jatkar:2009yd}), here we extend to F-theory.

The two main extensions considered here, namely singular Calabi--Yau threefolds and the ALE/ALF spaces transverse to the brane, lead to new 2d $\mathcal{N}=(0,4)$ SCFTs. We compute all relevant anomaly coefficients, i.e.~central charges and levels in both left- and right-moving sectors, using holography and supergravity techniques as introduced in \cite{Kraus:2005vz,Kraus:2005zm,Hansen:2006wu}. In particular we reduce the relevant 6d $\mathcal{N}=(1,0)$ supergravity theory on the compact part of the black strings. The anomaly coefficients are given by the coefficients of Chern--Simons terms in the resulting 3d action. Following \cite{Dabholkar:2010rm,Grimm:2018weo,Couzens:2019wls} we perform this reduction in the asymptotic geometry in order to include the contribution of degrees of freedom living outside of the horizon \cite{Banerjee:2009uk,Jatkar:2009yd}. This is the reason why the SCFTs  dual to the black strings probing ALE and ALF spaces have different anomaly coefficients, even though the near-horizon geometry in both cases is AdS$_3 \times \text{S}^3/\Gamma$. The so-called classical part of the anomaly coefficients is found by reducing the six-dimensional supergravity action to 3d and this computation is performed in section \ref{sec:classcontr}. For ALF transverse spaces there are additional contributions from one-loop Chern--Simons terms where massive Kaluza--Klein modes run in the loop. For the case of a smooth CY$_3$ and Taub-NUT as transverse space, these contributions were essential for the matching to the microscopic results \cite{Grimm:2018weo}.\footnote{Contributions of one-loop induced Chern--Simons terms to the entropy of black holes were also important in \cite{Couzens:2019wls,ArabiArdehali:2018mil,Hull:2020byc}.} We compute these quantum contributions for a general ALF transverse space and possibly singular CY$_3$ in section \ref{sec:quantumcontr}.

The plan for the paper is as follows. We begin by discussing the relevant 6d supergravity theories and black string solutions in section \ref{sec:setup}. We then compute the anomaly coefficients of the dual SCFTs in sections \ref{sec:classcontr} and \ref{sec:quantumcontr}. The full anomaly coefficients are presented in the summary section \ref{summary} and we end the paper in section \ref{sec:discu} with a discussion. In addition there are two appendices with some of the more technical details of the calculations.

\section{Macroscopic setup}\label{sec:setup}

Compactifying F-theory on an elliptically fibered Calabi--Yau threefold  CY$_3$ yields in the low energy limit a six-dimensional supergravity theory with chiral $\mathcal{N}=(1,0)$ supersymmetry. Depending on the choice of threefold, the six-dimensional theory contains a fixed number of massless hypermultiplets, vector multiplets and tensor multiplets. Within this theory it is possible to find black string solutions which are the supergravity description of self-dual strings charged under tensor multiplets. The tensor multiplets descend from expanding the RR four-form $C_4$ in harmonic forms on the base $B$ of the Calabi--Yau threefold and therefore the strings admit an interpretation as D3-branes compactified on a curve $C\subset B$ inside the base of the CY$_3$. On the worldsheet of these strings lives a field theory which flows in the IR to a 2d $\mathcal{N}=(0,4)$ SCFT. In this paper we will not attempt to construct the explicit 2d SCFTs, for recent work along these lines see \cite{Lawrie:2016axq,Hayashi:2019fsa}, instead we shall compute the central charges and current levels of these SCFTs using AdS/CFT. 

The strings we consider are embedded in an asymptotically $\mathbb{R} \times M_4\times \mathrm{S}^1$ spacetime, with the string wrapped on the S$^1$ and probing the transverse space $M_4$. Preservation of supersymmetry and the equations of motion imply that $M_4$ must be a hyper-K\"ahler manifold equipped with its Ricci-flat metric. In addition, we take it to be non-compact. Such four-dimensional spaces $M_4$ have been classified and fall into four categories depending on their asymptotic properties. These are the ALE, ALF, ALG and ALH spaces. At infinity the metrics approach a quotient of the flat metric on $\mathbb{R}^{4-k}\times \mathrm{T}^k$ with $k=0$ for ALE, $k=1$ for ALF and so forth. We will consider two of these classes in this paper, namely ALE and ALF transverse spaces.\footnote{No explicit metrics are known for ALG and ALH spaces.} A manifold with an ALE or ALF metric is diffeomorphic to $\mathbb{R}^+\times \mathrm{S}^3/\Gamma$, with $\Gamma$ a freely acting finite subgroup of $SU(2)$. The metric is ALE when it asymptotically approaches a quotient of the Euclidean flat space metric. It is ALF when it approaches the metric on $(\mathbb{R}^3 \times \mathrm{S}^1)/\Gamma$ at infinity. The ALE metrics are known to exist for all discrete subgroups $\Gamma \subset SU(2)$ and these admit an ADE-classification. ALF metrics are only known for the A-series and D-series in this classification. We have listed some important characteristic data of the groups $\Gamma$ and spaces $M_\Gamma$ in table \ref{Pontryagin table}. In the black string solutions we always take the spaces in the limit that they have an $\mathbb{R}^4/\Gamma$ singularity at their center. Despite this singularity the black string solutions have a smooth near-horizon limit. The final piece of the jigsaw necessary to understand the later computations is the isometry group of the various spaces. For the A-series both ALE and ALF spaces admit a $U(1)_{L}\times SU(2)_{R}$ isometry group, whilst for D- and E-series the isometry group is reduced to just $SU(2)_R$. 
\begin{table}[h]
\begin{center}
\begin{tabular}{lccc}
\toprule
$\Gamma\subset \text{SU}(2)$      \qquad          & $\qquad|\Gamma|$  &\qquad $p_1(M_\Gamma)$ (ALE) & \qquad $p_1(M_\Gamma)$ (ALF)  \\
\midrule
cyclic group $\mathbb{Z}_m$ \qquad   &\qquad$m$  &   \qquad $2m-\frac{2}{m}$ &\qquad   $2m$      \\[0.05cm]
binary dihedral  $\mathbb{D}^*_{m}$  \qquad  &\qquad$4m$  &   \qquad  $2m+6-\frac{1}{2m}$&  \qquad $2m+6$\\[0.05cm]
binary tetrahedral  $\mathbb{T}^*$    \qquad          &\qquad24 &\qquad $\frac{167}{12}$ &\qquad   \\[0.05cm]
binary octahedral  $\mathbb{O}^*$    \qquad         &\qquad48 &\qquad $\frac{383}{24}$ &\qquad \\[0.05cm]
binary icosahedral  $\mathbb{I}^*$      \qquad       &\qquad120  & \qquad $\frac{1079}{60}$ &\qquad \\
\bottomrule
\end{tabular}
\end{center}
\caption{The freely acting discrete subgroups of $SU(2)$, their sizes and the first Pontryagin numbers of the different ALE and ALF spaces $M_\Gamma$. The empty entries signify that such a space does not exist.} 
\label{Pontryagin table}
\end{table}

In the remainder of this section we first review 6d $\mathcal{N}=(1,0)$ supergravity obtained by compactifying F-theory on an elliptically fibered Calabi--Yau threefold in order to introduce the necessary notation and to set the scene. We then give the relevant black string solutions of this theory which will form the basis of our macroscopic computations.


\subsection{Six-dimensional $\mathcal{N}=(1,0)$ supergravity from F-theory on CY$_3$}\label{6d sugra}

In \cite{Bonetti:2011mw,Ferrara:1996wv} the authors reduced F-theory on an elliptically fibered Calabi--Yau threefold, which is not necessarily smooth, to obtain six-dimensional $\mathcal{N}=(1,0)$ supergravity. The multiplets of the theory are classified by representations of the little group $SU(2)\times SU(2)$ and are given in table \ref{Field content}.
\begin{table}[h]
\begin{center}
\begin{tabular}{m{2.0cm}m{3.7cm}m{4cm} m{2.2cm}<{\centering} m{1.1cm}<{\centering}}
\toprule
Multiplet    & 	Representation \newline little group   	& 		Field content 		&	  	Multiplicity & Gauge\newline rep.\\
\midrule
Gravity \newline multiplet & $(1,1) \oplus 2(\tfrac{1}{2}, 1) \oplus (1,0)$ &1 graviton\newline 1 Weyl LH gravitino\newline 1 self-dual 2-form& 1& $\boldsymbol{1}$\\

Tensor \newline multiplet & $(0,1)\oplus 2(0,\tfrac{1}{2})\oplus (0,0)$ & 1 anti-self-dual 2-form\newline 1 Weyl RH tensorino\newline 1 real scalar & $n_{T}$& $\boldsymbol{1}$\\

Hyper- \newline multiplet & $2(0,\tfrac{1}{2})\oplus 4 (0,0)$ & 1 Weyl RH hyperino\newline 2 complex scalars & $n_{H}$& $\RR$\\

Vector \newline multiplet & $(\tfrac{1}{2},\tfrac{1}{2})\oplus 2(\tfrac{1}{2},0)$ & 1 vector\newline 1 Weyl LH gaugino&$n_V$& $\boldsymbol{\mathrm{ad}}_{{i}}$\\
\bottomrule
\end{tabular}
\end{center}
\caption{Field content of 6d $\mathcal{N}=(1,0)$ supergravity. We have used the shorthand RH and LH for right- and left-handed.}
\label{Field content}
\end{table}
The multiplicities of the matter fields are not arbitrary and depend on the choice of Calabi--Yau threefold. The 6d spectrum also needs to satisfy anomaly constraints as we will briefly explain below. 

\begin{figure}[h!]
\begin{center}
\includegraphics[width=0.5\textwidth, angle=0]{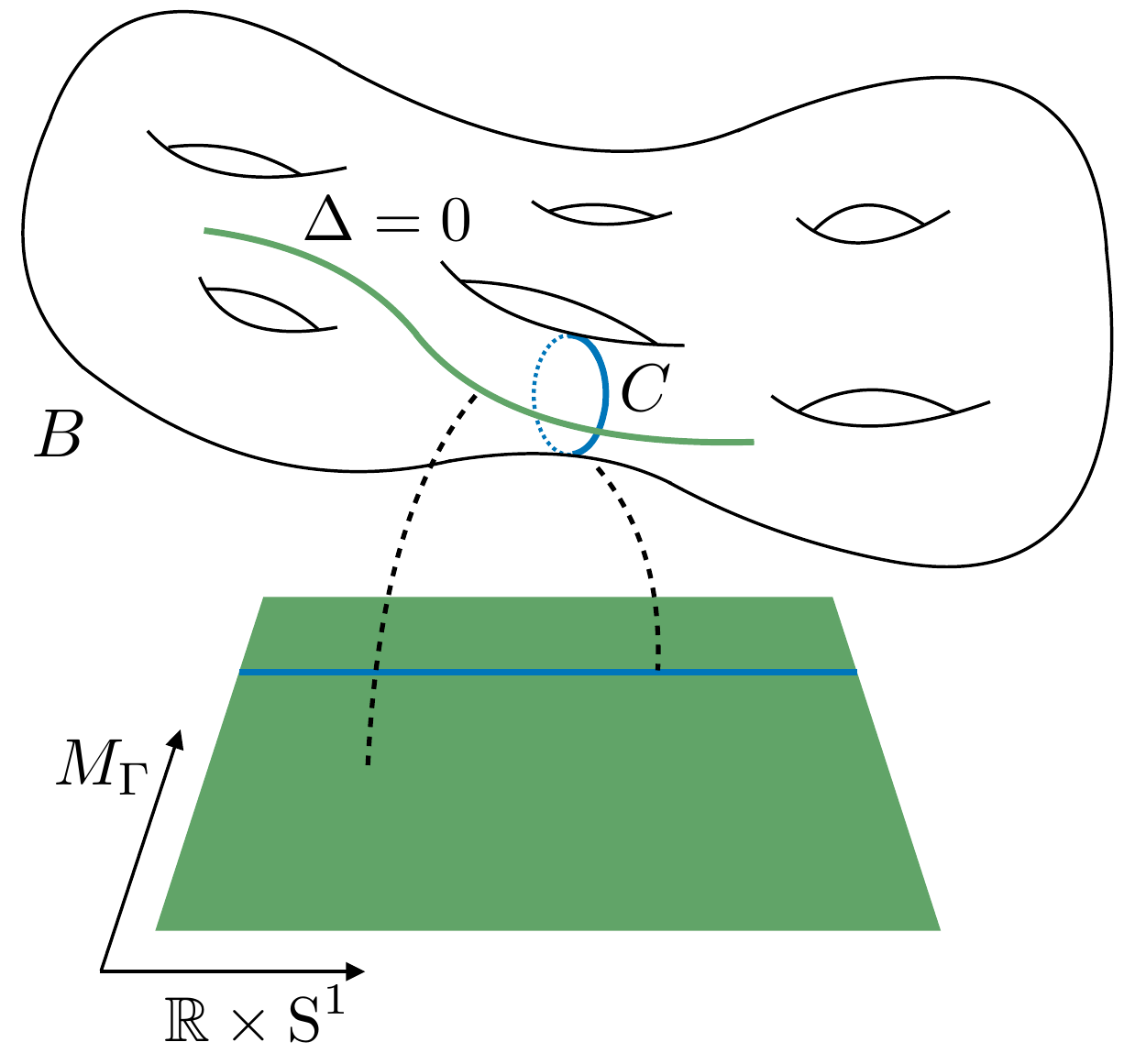}
\end{center}
\caption{The 7-branes (green) wrap the discriminant locus $\Delta=0 \subset B$ and extend along the dimensions $\mathbb{R} \times \text{S}^1 \times M_\Gamma$. The D3-brane (blue) wraps the curve $C \subset B$ and extends along $\mathbb{R} \times \text{S}^1$.}
\label{D3 wrapped pic}
\end{figure}

To classify the singularities of the elliptic fibration one can look at the vanishing of the discriminant $\Delta$ of the Weierstrass equation along the base of the Calabi--Yau. The complex codimension one loci determined by the vanishing of the discriminant signifies the loci of the 7-branes. The types of singular fibers, classified by Kodaira--N\'eron \cite{Kodaira,Neron}, characterize the singularities of the total space. In general there may be effective divisors $S_{i}$ over which the threefold develops a singularity, in the F-theory picture this is dual to the existence of a stack of multiple 7-branes\footnote{The fiber degenerations in the classification of Kodaira--N\'eron, which render the total space of the elliptic fibration smooth, are the type I$_{0,1}$ and type II fibers, see e.g.~table 4.1 in \cite{Weigand:2018rez}. F-theory compactifications on Calabi--Yau threefolds with these mild fiber degenerations lead to 6d $\cN=(1,0)$ supergravity theories without vector multiplets and charged matter.}. One can then expand the Poincar\'e dual of the discriminant as
\be\label{Discriminant}
[\Delta]= \sum_{i}  \nu_{i} [S_{i}]+ [\Delta']\,,
\ee
where $[\Delta']$ is associated to singularities of the fibration which render the total space smooth.\footnote{In the following we will drop the brackets to distinguish a curve and its Poincar\'e dual two-form, using the same symbol for both in slight abuse of notation.} The stacks of 7-branes along the divisors $S_i$ give rise to non-abelian gauge theories on their worldvolumes. A schematic illustration of the F-theory geometry and the wrapped branes therein is given in figure \ref{D3 wrapped pic}.

The exact nature of the gauge group is determined by the type of singularity of the total space and admits an ADE classification. These gauge groups naturally descend to become the gauge groups of the 6d theory. In the following we will assume that the gauge group $G$ is a product of simple factors,
\be
G=\prod_{i} G_{i}\, .
\ee
Each simple factor $G_i$ arises from a 7-brane stack along $S_i$ in the underlying F-theory geometry with associated singularity of the threefold. The gauge fields have the usual non-abelian field strength
\be
\hat{F}^{i}= \dd \hat A^{i}+ \hat A^{i}\wedge \hat A^{i}
\ee
with associated Chern--Simons three-form
\be\label{CS form}
\omega^{\text{CS}}(\hat A^{i})= \tr \Big( \hat A^i \wedge \dd\hat A^i +\frac{2}{3} \hat A^i \wedge \hat A^i \wedge \hat A^i\Big)\,.
\ee

Let us denote a basis of two-forms on the base of the elliptically fibered Calabi--Yau threefold by $\{ \omega_{\alpha}\}$ with $\alpha=1,\ldots, h^{1,1}(B)$. Associated to this basis is an intersection matrix 
\be\label{intmatrix}
\eta_{\alpha\beta}=\int_{B} \omega_{\alpha}\wedge \omega_{\beta}\,
\ee 
which defines an inner product for objects in $H^{1,1}(B)$. The matrix is of mostly minus Lorentzian signature and is  invariant under $SO(1,h^{1,1}(B)-1)$ transformations. We can expand the Poincar\'e dual of the divisor $S_{i}$ in terms of this basis as $S_{i}= s_{i}^{\alpha}\omega_{\alpha}$, with $s_{i}^{\alpha}$ constants. Likewise the Poincar\'e dual of the curve $C$ and the first Chern class of the base $c_1(B)$ can be expanded as $C=q^\alpha \omega_\alpha$ and $c_{1}(B)=c^{\alpha} \omega_{\alpha}$ respectively, with $q^\alpha$ and $c^{\alpha}$ constants. These constant coefficients will play a prominent role in the following sections. 
 
The scalars $\hat q^{U}$ of the hypermultiplets parametrize a quaternionic manifold. Since the explicit details of this manifold will not play a role in the following we shall not discuss it further and just denote the metric on this space by $h_{UV}$. The scalars are gauged with respect to the vector multiplets in the usual minimal coupling manner
\be
\mathcal{D} \hat q^{U}= \dd  \hat q^{U}+ \hat A^{i} (T_{i}^{\RR_i} \hat q)^{U}\, ,
\ee
where $T_i^{\RR_i}$ are the group generators acting on the scalars $\hat q^U$ in the representation $\RR_i$ of $G_i$. The scalars of the tensor multiplets parametrize the scalar manifold 
\be
\frac{SO(1,n_T)}{SO(n_T)}\, ,
\ee
and are uncharged under the gauge group $G$. To describe them it is convenient to introduce $n_T+1$ scalars $\hat \jmath^{\alpha}$, $\alpha=1,\ldots, n_T+1$, satisfying the constraint $\eta_{\alpha\beta} \hat \jmath^{\alpha} \hat \jmath^{\beta}=1$. Finally let us consider the two-forms in the theory. We will denote all the two-forms, i.e.~two-forms in the gravity and tensor multiplets, collectively as $\hat B^{\alpha}$. The gauge invariant field strength of the tensors is given by
\be
\hat{G}^{\alpha}=\dd \hat{B}^{\alpha}+\frac{1}{2} c^{\alpha} \hat{\omega}^{\mathrm{CS}}_{\text{grav}}-2 s_{i}^{\alpha} \omega^{\mathrm{CS}}(\hat A^{i})\, ,
\ee
with the $s_{i}^{\alpha}$ and $c^{\alpha}$ as introduced below \eqref{intmatrix}. Using the scalars $\hat \jmath^\alpha$ we can define the positive definite metric
\be
g_{\alpha \beta}=2\hat \jmath_\alpha \hat \jmath_\beta-\eta_{\alpha \beta}\,.
\ee 
such that the (anti-)self-duality constraints satisfied by the two-forms can be collectively written as
\be\label{self duality}
g_{\alpha \beta}\hat  \ast \hat{G}^{\beta}= \eta_{\alpha\beta}\hat{G}^{\beta}\, .
\ee
Due to these (anti-)self-duality constraints it is strictly speaking only possible to construct a pseudo-action. This will not pose a problem so long as we correctly implement the self-duality constraints at the level of the equations of motion. The bosonic action of gauged $\mathcal{N}=(1,0)$ supergravity is given by\footnote{We use conventions where $\kappa_6^2=(2\pi)^3$.}
\begin{align}
S^{(6)}= \frac{1}{(2\pi)^3}\int_{M_6} &\bigg[\frac{1}{2} \hat{R}\hat \ast 1- h_{UV} \mathcal{D} \hat q^U \wedge \hat \ast \mathcal{D} \hat q^V-\frac{1}{4} g_{\alpha\beta} \hat{G}^{\alpha} \wedge \hat{\ast}\hat{G}^{\beta}-\frac{1}{2} g_{\alpha\beta} \dd \hat \jmath^{\alpha} \wedge \hat \ast \dd \hat \jmath^{\beta}\nonumber\\
&+2 \sum_{i}  \eta_{\alpha\beta} \hat \jmath^{\alpha} s^{\beta}_{i} \tr \hat{F}^{i} \wedge \hat{\ast} \hat{F}^{i}- \frac{1}{4} \eta_{\alpha\beta} \hat{B}^{\alpha} \wedge \hat{X}_{4}^{\beta} - \mathcal{V} \hat{\ast}1\bigg]\, .\label{action}
\end{align}
The precise form of the potential does not concern us, though it may be found in \cite{Suzuki:2005vu} for example. 
The Green--Schwarz--Sagnotti--West coupling $\hat{X}_{4}^{\alpha}$ is given by 
\begin{align}
\hat{X}_{4}^{\alpha}&=\dd \hat{X}_{3}^{\alpha} =\frac{1}{2} c^\alpha \tr \hat{\mathcal{R}}\wedge \hat{\mathcal{R}}-2 \sum_{i} s_{i}^{\alpha} \tr \hat{F}^{i}\wedge \hat{F}^{i}\, ,\nn\\
\hat{X}_{3}^{\alpha}&=\frac{1}{2} c^{\alpha} \hat \omega^{\text{CS}}_{\text{grav}} -2 \sum_{i} s_{i}^{\alpha} \omega^{\text{CS}}(\hat A^{i})\, .
\end{align}
Here the gravitational Chern--Simons term takes the usual form
\be\label{CSgrav}
\hat \omega^{\text{CS}}_{\text{grav}}=\tr \Big(\omega \wedge \dd \omega +\frac{2}{3}\omega \wedge \omega \wedge \omega\Big)\, ,
\ee
with $\omega$ the spin connection in 6d. 

Since we consider six-dimensional theories which have a string embedding we must impose a set of necessary consistency conditions for the theory to be anomaly free.\footnote{However, a six-dimensional theory satisfying these conditions may still be in the swampland.} With the aid of the concrete string embedding, these conditions may be written in terms of topological quantities of the underlying compactification geometry. Let us first introduce some notation. Denote by $\tr_{\RR_i}$  the trace in the representation $\RR_i$ of $G_i$, and let $\boldsymbol{\mathrm{f}}_i$ denote the fundamental representation of $G_i$. We define the normalization constants $\lambda_i$ which depend on the simple group factor and whose purpose is such that the trace operator for the field strengths appearing above satisfies $\lambda_i \tr =\tr_{\boldsymbol{\mathrm{f}}_{i}}$. The normalization constant is fixed by requiring the lowest topological charge of an $SU(2)$ instanton to be 1 with respect to the trace $\tr$.\footnote{The values of the constants can be found in \cite{Kumar:2010ru} among other places.} Next, define the group theory coefficients  $A_{\RR_i} ,\, B_{\RR_i}, \, C_{\RR_i}$ via
\begin{align}
\tr_{\RR_i} \hat {F}^{i \,2}&=A_{\RR_i} \tr_{\boldsymbol{\mathrm{f}}_i} \hat {F}^{i \,2}\, ,\nn\\
\tr_{\RR_i} \hat F^{i \,4}&= B_{\RR_i} \tr_{\boldsymbol{\mathrm{f}}_i} \hat F^{i \,4}+ C_{\RR_i}(\tr_{\boldsymbol{\mathrm{f}}_i} \hat F^{i \,2})^2\, .
\end{align}
Further, let the multiplicity of the irreducible representation $\RR_i$ of $G_i$ be $x_{\RR_i}^{i}$, and the multiplicity of the irreducible representation $(\RR_i,\RR_j)$ of $G_i\times G_j$ be $x_{\RR_i\RR_j}^{ij}$. Then the anomaly constraints read
\begin{align}
n_{H}-n_{V}&=273-29 n_{T}\, ,\label{grav anomaly}\\
n_T&=9 - \eta_{\alpha \beta }c^\alpha c^\beta\, ,\label{nT}\\
B_{\boldsymbol{\mathrm{ad}}_{i}}&=\sum_{\RR_i} x^{i}_{\RR_i} B_{\RR_i}\, ,\\
 \eta_{\alpha\beta} c^{\alpha}s^{\beta}_{i}&= \frac{1}{6} \lambda_{i}\bigg( \sum_{\RR_i} x^{i}_{\RR_i}A_{\RR_i}- A_{\boldsymbol{\mathrm{ad}}_{i}}\bigg)\, , \label{c.s}\\
\eta_{\alpha\beta} s_{i}^{\alpha} s_{i}^{\beta}&=\frac{1}{3} \lambda_{i}^2 \bigg(\sum_{\RR_i} x^{i}_{\RR_i}C_{\RR_i}-C_{\boldsymbol{\mathrm{ad}}_{i}}\bigg)\, , \qquad \text{no sum over}~i \\
\eta_{\alpha\beta} s^{\alpha}_{i}s^{\beta}_{j}&= \lambda_{i}\lambda_{j} \sum_{\RR_i,\RR_j} x^{ij}_{\RR_i\RR_j} A_{\RR_i}A_{\RR_j}\, , \qquad \quad\,(i \neq j)\, .
\end{align}
In order for the theory to be anomaly free each of the above conditions must be satisfied.


\subsection{Black string solutions}\label{black string}

In \cite{Lam:2018jln,Grimm:2018weo} black string solutions of two-derivative $\mathcal{N}=(1,0)$ supergravity coupled to an arbitrary number of tensor multiplets were constructed. These solutions can also be embedded in the supergravity theory containing hyper- and vector multiplets, as given in the previous section, by setting the hyperscalars to constants and the vector fields to zero. Clearly this solution essentially truncates away the hyper- and vector multiplets, nevertheless when computing the levels these multiplets will still play a role. The metric for the black string is 
\be\label{blackstringsolution}
\dd \hat{s}_6^2= 2 H^{-1}(\dd u+\beta)\Big (\dd v +\omega +\frac{1}{2} \mathcal{F}(\dd u+\beta) \Big) + H \dd s^2 (M_\Gamma)\, ,
\ee
with $\dd s^{2}(M_\Gamma)$ the Ricci-flat metric for the ALE or ALF space with ADE subgroup $\Gamma$. The solution is supported by a non-trivial three-form field strength. The scalar functions $\mathcal{F}$ and $H$ and one-forms $\beta$ and $\omega$ are independent of the $u$ and $v$ coordinates and can be derived by solving certain differential equations given in \cite{Lam:2018jln} once the explicit metric on $M_\Gamma$ has been chosen. When the Killing vector $\partial_{u}$ is taken to be spacelike, the metric corresponds to a black string wound in the $u$-direction. 

The near-horizon limit of this geometry is fixed to be AdS$_3\times \mathrm{S}^3/\Gamma$ \cite{Couzens:2017way} as expected from our comments about the existence of a 2d SCFT living on the worldvolume of the strings and consistency with AdS/CFT. Note that the near-horizon geometry is blind to the full details of its UV origin: it is unable to distinguish between an ALE or ALF UV completion. The asymptotic form of the solution is easily constructed by replacing the metric on $M_\Gamma$ by its asymptotic metric $\dd s^2(M_{\Gamma}^{(\infty)})$. For ALE and ALF spaces this is given by a $\Gamma$ quotient of the covering space, which we denote by $M^{(\infty)}$, and whose asymptotic metric is
\be\label{covering space}
\dd s^2 (M^{(\infty)})=
\begin{cases}
\dd r^2+\frac{1}{4} r^2 (\sigma_{1}^2+\sigma_2^2+\sigma_3^2)\,,\quad M_{\Gamma} \text{ is ALE}\,,\\
\dd r^2+ r^2 (\sigma_1^2+\sigma_2^2)+\sigma_3^2\,, \quad \; \; M_{\Gamma} \text{ is ALF}\,,
\end{cases}
\ee
with $\sigma_i$ the left-invariant Maurer--Cartan forms on S$^3$. The usefulness of the covering space is that it can be used to treat all quotients simultaneously. 

The charges of the black string are obtained by integrating the field strengths $\hat G^{\alpha}_{\Gamma}$ over the compact part of the geometry of the quotient space, given by the $\sigma_{i}$ part in equation \eqref{covering space}, and denoted by $M_{\Gamma}^{\text{sph}}$ from now on. We will use the same conventions as in \cite{Grimm:2018weo,Couzens:2019wls} for normalizing the charge, i.e.
\be
\int_{M_{\Gamma}^{\text{sph}}} \hat{G}^{\alpha}_{\Gamma}=\frac{1}{|\Gamma|} \int_{M^{\text{sph}}}\hat{G}^{\alpha}=-(2 \pi)^2 Q^{\alpha}~.
\ee
In going from the first expression to the second we have used the quotient map ${\pi: M\rightarrow M_{\Gamma}}$, which acts on the three-forms as $\hat{G}^{\alpha} = \pi^{*} \hat{G}_{\Gamma}^{\alpha}$. To relate these macroscopic charges to their microscopic counterparts $q^\alpha$ in the expansion $C=q^\alpha \omega_\alpha$ we use the results of \cite{Grimm:2018weo}. There it was shown that 
\be\label{Charge shift}
Q^{\alpha}=q^{\alpha}-\frac{1}{4} p_{1}(M_{\Gamma}) c^{\alpha}
\ee
with $p_{1}(M_{\Gamma})$ the first Pontryagin number of $M_{\Gamma}$ as given in table \ref{Pontryagin table}\footnote{They can be computed using the data in \cite{Gibbons:1979gd}.} and the $c^{\alpha}$ as given below equation \eqref{intmatrix}. The charge shift is due to the
\be
\eta_{\alpha\beta} c^{\alpha}\hat{B}^{\beta}\wedge \tr \hat{\mathcal{R}}\wedge\hat{\mathcal{R}}
\ee 
term in \eqref{action}.\footnote{When the elliptic fibration is trivial, and we are thus in type IIB on K3 or T$^4$, there are no such higher derivative corrections which is why there is no charge shift in \cite{Couzens:2019wls}.} When integrating the tensor field equation of motion over the transverse space $M_{\Gamma}$ in the presence of string-like sources with microscopic charges $q^{\alpha}$, this term generates a shift of the charge by a term proportional to
\be
\int_{M_{\Gamma}} \tr \hat{\mathcal{R}}\wedge\hat{\mathcal{R}} \propto p_1(M_\Gamma)\,.
\ee
One may wonder if there is a further charge shift due to the gauge fields and the coupling
\be
\eta_{\alpha\beta}s^{\alpha}_{i} \hat B^{\beta}\wedge \tr \hat{F}^{i}\wedge \hat{F}^{i}\, 
\ee
appearing in the action. Indeed, if in the background of the black string vector multiplets are turned on with field strengths satisfying
\be
\int_{M_{\Gamma}}\tr \hat{F}^{i}\wedge \hat{F}^{i}\neq0
\ee
this would induce a further shift in the charge. From the F-theory perspective these instantons can be viewed as gauge instantons living on the 7-branes. The instantons induce D3-brane charge localized on the 7-brane worldvolume transverse to the instanton configuration through a Chern--Simons coupling of the RR four-form and the gauge fields.\footnote{This is similar to D(-1)-brane charge induced on D3-branes via gauge instantons in $\mathcal N=4$ SYM. In the case of the 7-branes, the coupling in question has the form $C_4 \wedge \mathrm{tr}\, F \wedge F$.} We will consider solutions where such instanton configurations are not present and therefore the dictionary between macroscopic and microscopic quantities is given by \eqref{Charge shift}. It may be interesting in the future to study such instanton configurations. For Taub-NUT space instanton configurations have been discussed in \cite{Cherkis:2009jm,Witten:2009xu,Cherkis:2016gmo}.


\section{Classical contributions}\label{sec:classcontr}

In this section we compute the classical contribution to the central charges and levels of current algebras of the black string solutions.\footnote{We call these the classical contributions since they are computed by dimensionally reducing the six-dimensional action to three dimensions. In the next section we will consider the contributions from integrating out massive KK modes which are quantum in nature.} One can then use these results to make a prediction about the corresponding microscopic anomaly polynomial of the dual 2d $\mathcal{N}=(0,4)$ SCFTs, and compare for example with the prediction one can infer from the conjecture of \cite{Bah:2020jas}. 

In both the ALE and ALF cases the near-horizon geometry takes the product form AdS$_3\times \mathrm{S}^3/\Gamma$. In contrast the asymptotic regime of the two cases does not suffer this degeneracy. Both cases approach $\mathbb{R}^{1,1}\times M_{\Gamma}^{(\infty)}$ with $M_{\Gamma}^{(\infty)}$ given by a quotient of \eqref{covering space}. Naively one would compute the central charges and current levels of the 2d SCFT by dimensionally reducing the six-dimensional action on the compact space in the near-horizon to obtain an AdS$_3$ effective action \cite{Kraus:2005vz,Kraus:2005zm,Hansen:2006wu}. The central charges and current levels can then be read off from the resultant Chern--Simons terms using the AdS/CFT dictionary. However in \cite{Dabholkar:2010rm} and further confirmed in \cite{Grimm:2018weo, Couzens:2019wls} this was shown to be insufficient to reproduce the correct microscopic results. A quick way of seeing that this is not the correct procedure is to observe that there is nothing to distinguish between a D3-brane probing an ALE or ALF space via this method of looking at the near-horizon geometry. This near-horizon analysis precisely misses the degrees of freedom living outside the near-horizon region which also contribute to the microscopic degeneracy \cite{Banerjee:2009uk,Jatkar:2009yd}.

Instead the correct procedure is to reduce the action on the compact part of the asymptotic geometry \cite{Grimm:2018weo, Couzens:2019wls}. Again this is not quite the full story, one must also integrate out massive Kaluza--Klein (KK) modes and include their contributions to the central charges and levels. We refer to the contributions arising from the massive KK modes as quantum contributions and to those from the reduction of the action as classical. Computing these classical contributions will be the content of this section, whilst the computation of the quantum corrections will be the subject of the subsequent section. 

We split the discussion into two parts. In the first part we compute the levels $k_{L,R}$ associated to the current algebras of 
the $U(1)_L$ (when it exists) and the $SU(2)_R$. In addition we compute the difference of the central charges $c_L-c_R$ which  follows from 
the coefficient of the 3d gravitational Chern--Simons term. We refer to $k_{L,R}$ as the universal levels in the 
following. When $SU(2)_R$ can be identified with the right-moving $SU(2)_r$ R-symmetry, supersymmetry implies 
that the right-moving central charge and universal level are related by $c_R=6k_R$ and both the central charges can 
be determined.\footnote{However, the identification of $SU(2)_R$ with the R-symmetry might fail for (a small) part of the spectrum, e.g. the 
center of mass modes, and the result for $c_{L,R}$ determined this way might differ from the central charges at 
subleading order in the charges, see e.g. \cite{Dabholkar:2010rm,Couzens:2019wls}.} In the second part of this section we compute the levels of non-abelian flavour symmetries 
associated to 6d vector multiplets coming from the reduction on singular Calabi--Yau threefolds.

\subsection{Central charges and universal levels}

Let us now proceed in determining the classical contribution to the central charges and levels associated to the current algebras arising from the isometries of the solution. Recall that for the A-series we have a $U(1)_L\times SU(2)_R$ isometry group whilst for D- and E-series the isometry group is reduced to $SU(2)_R$. We will reduce the six-dimensional pseudo-action on the compact space of the black string solution at asymptotic infinity in order to obtain Chern--Simons terms for the associated current algebra. Since the reduction works similarly for the ALE and ALF transverse spaces we shall perform the reduction simultaneously. This can be done via the covering space at asymptotic infinity which is given by the black string solution \eqref{blackstringsolution} with $\mathrm{d}s^2(M_\Gamma)$ replaced by
\begin{equation}
\mathrm{d}s^2_4=V^{-1}(\mathrm{d}\psi+\cos(\theta)\mathrm{d}\phi)^2+V\big(\mathrm{d}r^2+r^2\mathrm{d}\theta^2+r^2\sin^2(\theta)\mathrm{d}\phi^2\big)\,\label{coveringspace}
\end{equation}
in the limit that $r\rightarrow \infty$. Here $0\leq \psi<4\pi$, $0\leq \phi < 2\pi$, $r>0$, $0\leq \theta <\pi$ and 
\begin{equation}
V=v_\infty+\frac{1}{r}\,,\qquad   \begin{cases}v_\infty=0\quad \text{for ALE}\,, \\ v_\infty \neq 0 \quad \text{for ALF}\,. \end{cases}
\end{equation}
Indeed in the limit $r\rightarrow \infty$ this agrees with the asymptotic metric \eqref{covering space}. One proceeds by gauging the isometries of the compact part of the solution, i.e. gauge the symmetries acting on the $SU(2)$ Maurer--Cartan one-forms. We are therefore gauging the isometries of S$^3/\Gamma$.

We split the ansatz for the covering space as
\be
\dd \hat{s}^{2}_{6}=\dd s^2_{M_3}+ \delta_{ab} e^{a} e^{b}
\ee
with $M_3$ the three-dimensional non-compact part of spacetime, and
\be\label{vielbeinans}
e^{a} =\begin{cases} 
\hat{e}^{a}- K^{Ia}_{R} A^{I}_{R}- K^{a}_{L} A_{L} &\quad \text{A-series}\,,\vspace{.2cm}\\
\hat{e}^{a}- K^{Ia}_{R} A^{I}_{R}& \quad \text{D- and E-series} \, .
\end{cases}
\ee
Here $A_L$, $K_L$, $A_R$ and $K_R$ are the gauge fields and Killing vectors corresponding to the isometry group of the transverse space. Explicit expressions for the Killing vectors can be found in \cite{Grimm:2018weo}. The dreibein $\hat{e}^{a}$ corresponds to the spherical part of the black string solution \eqref{blackstringsolution} with base space \eqref{coveringspace} and are given by 
\begin{align}
\hat e^1&=\sqrt{HV}r\big(\sin(\psi)\mathrm{d}\theta-\cos(\psi)\sin{\theta}\mathrm{d}\phi\big)\,,\nn \\
\hat e^2&=\sqrt{HV}r\big(\cos(\psi)\mathrm{d}\theta+\sin(\psi)\sin{\theta}\mathrm{d}\phi\big)\,,\\
\hat e^3&=\sqrt{H/V}\big(\mathrm{d}\psi+\cos(\theta)\mathrm{d}\phi\big)\,.\nn
\end{align}
For the three-form $\hat{G}^{\alpha}$ on the covering space we take as ansatz
\be\label{ansatz three-form}
\hat{G}^{\alpha}=-Q^{\alpha} [(2\pi)^2 |\Gamma|(e_3-\chi_3)+\omega(M_3)]\, , 
\ee
where $\omega(M_3)$ is a three-form on $M_3$ which is necessary for the existence of the solution but whose explicit form is not required. The three-form $e_3$ is the invariant volume form on the gauged compact space. It is normalized such that
\be
\int_{M^{\text{sph}}} e_3=1\, 
\ee
and it has exterior derivative
\be
\dd e_3=\begin{dcases}
\frac{1}{8\pi^2} \tr F_R\wedge F_R+\frac{1}{16 \pi^2}F_L\wedge F_L \qquad & \text{A-series}\,,\vspace{.2cm}\\[0.2cm]
\frac{1}{8 \pi^2} \tr F_R \wedge F_R\qquad & \text{D- and E-series}\,.
\end{dcases} \, 
\ee
Its explicit form can be found in appendix \ref{appendixredhigh}.
Finally, the three-form $\chi_3$ is included in order for the ansatz to satisfy its Bianchi identity and takes the form
\be\label{chi3}
\chi_{3}=\begin{dcases}
\frac{1}{8\pi^{2}}\mathrm{tr}\Big(A_{R}\wedge\mathrm{d}A_{R}+\frac{2}{3}A_{R}^{3}\Big)+\frac{1}{16 \pi^2} A_{L}\wedge F_{L} \quad & \text{A-series}\,,\vspace{.2cm}\\[0.2cm]
\frac{1}{8\pi^{2}}\mathrm{tr}\Big(A_{R}\wedge\mathrm{d}A_{R}+\frac{2}{3}A_{R}^{3}\Big)\quad & \text{D- and E-series}\,.
\end{dcases} 
\ee
Having introduced the necessary notation and conventions we can proceed with computing the difference of the central charges and the levels of the various universal current algebras. These are related to the following Chern--Simons terms in the 3d effective action 
\be\label{Gen CS}
S_{\text{CS}}\supset \frac{k_{L}}{8\pi} \int_{M_{3}} A_{L}\wedge F_{L} +\frac{k_{R}}{4\pi}\int_{M_3}\omega^{\mathrm{CS}}(A_{R})+ \frac{c_{L}-c_{R}}{96 \pi} \int_{M_3} \omega^{\mathrm{CS}}_{\text{grav}}\, ,
\ee
where it is understood that for the D- and E-series the first term is absent. The only part of the 6d action which yields such Chern--Simons terms is given by 
\be
S^{(6)}\supset \fr{1}{(2\pi)^3}\int_{M_6} \big[ -\tfr{1}{4} \,  g_{\ax \bx}\,  \hat G^\ax_\Gamma \wedge  \hat  \ast \, \hat G^\bx_\Gamma-\tfr{1}{8}\eta_{\ax \bx} \,c^\ax \hat B^\bx_\Gamma \wedge  \tr \hat \cR \w \hat \cR \big]\, ,\lab{termscontributing}
\ee
and we restrict our focus to this part of the action in the following section. We compute the contributions arising from two- and four-derivative terms in \eqref{termscontributing} separately.


\subsubsection{Two-derivative contributions}

To determine the contributions from the two-derivative part of the action we perform a gauge transformation for the universal current algebras $\mathfrak{u}(1)_L\oplus \mathfrak{su}(2)_R$ for the A-series and $\mathfrak{su}(2)_R$ for the D- and E-series. We then reduce this variation on the spherical part of the geometry and relate the result to the variation of the 3d Chern--Simons terms \eqref{Gen CS}. We denote the gauge parameter by $\Lambda$ in the following. The three-form $e_3$ is gauge invariant by construction and therefore the only contribution to the variation of the two-derivative terms in \eqref{termscontributing} will arise from varying $\chi_3$. Performing the above steps, and using the covering space to compute the integral, we find\footnote{We use conventions in which $\int_{M_6}=\int_{{M}_3}\cdot \int_{M_\Gamma^{\mathrm{sph}}}$.}
\begin{align}\label{2 deriv}
\delta_{\Lambda} \mathcal{L}^{2\text{-der}}_{\text{CS}} \ast_3 1&= -\frac{1}{16\pi^3} \int_{M_{\Gamma}^{\text{sph}}} g_{\alpha \beta} \delta_{\Lambda}\hat{G}_{\Gamma}^{\alpha} \wedge \hat{\ast} \hat{G}_{\Gamma}^{\beta}=-\frac{1}{16\pi^3|\Gamma|} \int_{M^{\text{sph}}} g_{\alpha \beta} \delta_{\Lambda}\hat{G}^{\alpha} \wedge \hat{\ast} \hat{G}^{\beta}\nonumber\\
&=\pi|\Gamma| \eta_{\alpha \beta} Q^{\alpha} Q^{\beta}\int_{M^{\text{sph}}}\delta_{\Lambda}\chi_3\wedge e_3=\pi|\Gamma| \eta_{\alpha \beta} Q^{\alpha} Q^{\beta}\delta_{\Lambda}\chi_3\, .
\end{align}
In going from the first line to the second line we have used the self-duality constraints \eqref{self duality}. By comparison with the variation of \eqref{Gen CS} we obtain
\begin{align}
k_{L}^{2\text{-der}}&=\frac{1}{2} |\Gamma| \eta_{\alpha\beta} Q^{\alpha}Q^{\beta}\, ,\nonumber \\
k_{R}^{2\text{-der}}&=\frac{1}{2} |\Gamma| \eta_{\alpha\beta} Q^{\alpha}Q^{\beta}\, ,\label{2deriv}\\
c_{L}^{2\text{-der}}&= c_{R}^{2\text{-der}}\, , \nonumber
\end{align}
where $Q^{\alpha}$ is defined as in \eqref{Charge shift} and $k_{L}$ is only present for the A-series. The central charges are equal because the reduction of the two-derivative term in \eqref{termscontributing} does not generate a 3d gravitational Chern--Simons term.


\subsubsection{Four-derivative contributions}

Let us now consider the contributions from the four-derivative term in \eqref{termscontributing}. We can rewrite this term in terms of objects on the covering space of the solution as
\begin{equation}\label{higherderterm1}
 \frac{1}{64 \pi^2}\int_{M_{\Gamma}^{\text{sph}}} \eta_{\alpha \beta} c^{\alpha} \hat{G}^{\beta}_{\Gamma}\wedge \hat{\omega}_{\text{grav},\Gamma}^{\text{CS}}
=\frac{1}{64 \pi^2|\Gamma|}\int_{M^{\text{sph}}} \eta_{\alpha \beta} c^{\alpha} \hat{G}^{\beta}\wedge \hat \omega_{\text{grav}}^{\text{CS}}\, ,
\end{equation}
where $\hat{\omega}^{\text{CS}}_{\text{grav}, \Gamma}$ is the 6d gravitational Chern--Simons term of the $\Gamma$ quotiented space. We evaluate \eqref{higherderterm1} for the ansatz with respect to the black string solution \eqref{blackstringsolution} with base space \eqref{coveringspace} and will only take the $r\rightarrow \infty$ limit afterwards. In the process we only keep terms leading to 3d Chern--Simons terms. The details of this computation can be found in appendix \ref{appendixredhigh} but the result is given by
\begin{align}\label{resulthigher}
\mathcal{L}_{\text{CS}}^{4\text{-der}}\ast_3 1&=  \frac{1}{16} \eta_{\alpha \beta} c^{\alpha}Q^{\beta}\Biggl[\omega_{\mathrm{grav}}^{\mathrm{CS}}-\frac{1+4v_\infty r+2v_\infty^2 r^2}{(1+v_\infty r)^4}A_{L}\wedge F_{L} \\
  & \qquad  \qquad \qquad \qquad+2\frac{1+4v_\infty r+10v_\infty^2 r^{2}+8v_\infty^3 r^{3}+2v_\infty^4 r^{4}}{(1+v_\infty r)^{4}}\omega^{\mathrm{CS}}(A_{R})\Biggl]\,,\nn
\end{align}
where one has to set $A_L=0$ for the D- and E-series.  By setting $v_\infty=0$ for ALE transverse spaces we find
\be
\mathcal{L}_{\text{CS}}^{4\text{-der}}\ast_3 1=
\begin{dcases}
\frac{1}{16 \pi} \eta_{\alpha \beta} c^{\alpha} Q^{\beta} \big[ \omega_{\text{grav}}^{\text{CS}}- A_{L}\wedge F_{L} +2 \omega^{\text{CS}}(A_{R})\big]\, \quad &\text{A-series}\,,\vspace{0.2cm}\\[0.2cm]
\frac{1}{16 \pi} \eta_{\alpha \beta} c^{\alpha} Q^{\beta} \big[ \omega_{\text{grav}}^{\text{CS}}+2 \omega^{\text{CS}}(A_{R})\big]\, \quad &\text{D- and E-series}\,.
\end{dcases}
\ee
Comparing this to \eqref{Gen CS} we find that the four-derivative part of the central charges and levels is given by
\begin{align}
k_{L}^{4\text{-der}}&= -\frac{1}{2} \eta_{\alpha \beta}c^{\alpha} Q^{\beta}\, \qquad \text{A-series only}\,, \nonumber \\
k_{R}^{4\text{-der}}&= \frac{1}{2} \eta_{\alpha \beta} c^{\alpha} Q^{\beta}\, ,\label{ALE4deriv}\\
(c_L-c_R)^{4\text{-der}}&= 6 \eta_{\alpha\beta} c^{\alpha} Q^{\beta}\, .\nonumber
\end{align}
The Chern--Simons terms for ALF transverse spaces are obtained by taking $v_\infty \neq 0$ and taking the limit $r\rightarrow \infty$ in \eqref{resulthigher}:
\be
\mathcal{L}_{\text{CS}}^{4\text{-der}}\ast_3 1=\frac{1}{16 \pi}\eta_{\alpha\beta} c^{\alpha} Q^{\beta}\big[\omega_{\text{grav}}^{\text{CS}} +4 \omega^{\text{CS}}(A_{R})\big]\, .
\ee
The four-derivative part of the central charges and levels is thus equal to
\begin{align}
k_{L}^{4\text{-der}}&=0\, , \nonumber\\
k_{R}^{4\text{-der}}&= \eta_{\alpha \beta} c^{\alpha} Q^{\beta}\, ,\label{ALF4deriv}\\
(c_L-c_R)^{4\text{-der}}&= 6 \eta_{\alpha\beta} c^{\alpha} Q^{\beta}\, .\nonumber
\end{align}
The computation of the contribution to the left level is of course only relevant for the A-series.

\subsubsection{Total classical contributions}

We can now compute the total classical contribution by simply adding the result in \eqref{2deriv} to \eqref{ALE4deriv} for ALE spaces and to \eqref{ALF4deriv} for ALF spaces. We use that the macroscopic charges are related to the microscopic charges via \eqref{Charge shift}. For ALE transverse spaces we thus find
\begin{align}
k_{L}^{\text{class}}&= \frac{1}{2}|\Gamma| \eta_{\alpha\beta} Q^{\alpha}Q^{\beta}-\frac{1}{2} \eta_{\alpha\beta}c^{\alpha} Q^{\beta}\, \qquad \qquad \qquad \qquad \qquad \qquad \text{A-series only}\, \nn \\
&\equiv \frac{1}{2}|\Gamma|\Big( C -\frac{1}{4} p_{1}(M_{\Gamma}) c_{1}(B)\Big)^2-\frac{1}{2} c_{1}(B)\cdot\Big( C -\frac{1}{4} p_{1}(M_{\Gamma}) c_{1}(B)\Big)\, ,\nonumber \\
k_{R}^{\text{class}}&= \frac{1}{2} |\Gamma| \eta_{\alpha\beta} Q^{\alpha}Q^{\beta}+ \frac{1}{2} \eta_{\alpha\beta}c^{\alpha} Q^{\beta}\label{ALEclassical}\\%
&\equiv \frac{1}{2}|\Gamma|\Big( C -\frac{1}{4} p_{1}(M_{\Gamma}) c_{1}(B)\Big)^2+\frac{1}{2} c_{1}(B)\cdot\Big( C -\frac{1}{4} p_{1}(M_{\Gamma}) c_{1}(B)\Big)\, ,\nn\\
(c_{L}-c_{R})^{\text{class}}&=6 \eta_{\alpha\beta} c^{\alpha}Q^{\beta}\equiv 6 c_{1}(B)\cdot \Big( C -\frac{1}{4} p_{1}(M_{\Gamma}) c_{1}(B)\Big)\, .\nn 
\end{align}
For ALF transverse spaces the total classical contribution is given by
\begin{align}
k_{L}^{\text{class}}&= \frac{1}{2}|\Gamma| \eta_{\alpha \beta}Q^{\alpha}Q^{\beta}\,  \qquad \qquad \qquad \qquad \qquad \qquad \qquad \qquad \quad \text{A-series only}\, \nonumber \\
&\equiv \frac{1}{2}|\Gamma|\Big( C -\frac{1}{4} p_{1}(M_{\Gamma}) c_{1}(B)\Big)^2\, ,\nonumber \\
k_{R}^{\text{class}}&= \frac{1}{2} |\Gamma| \eta_{\alpha\beta} Q^{\alpha}Q^{\beta}+\eta_{\alpha\beta}c^{\alpha} Q^{\beta}\, \label{ALFclassical}\\
&\equiv \frac{1}{2}|\Gamma|\Big( C -\frac{1}{4} p_{1}(M_{\Gamma}) c_{1}(B)\Big)^2+ c_{1}(B)\cdot\Big( C -\frac{1}{4} p_{1}(M_{\Gamma}) c_{1}(B)\Big)\, ,\nn\\\
(c_{L}-c_{R})^{\text{class}}&=6 \eta_{\alpha\beta} c^{\alpha}Q^{\beta}\equiv 6 c_{1}(B)\cdot \Big( C -\frac{1}{4} p_{1}(M_{\Gamma}) c_{1}(B)\Big)\, .\nn
\end{align}
The first Pontryagin numbers $p_1(M_\Gamma)$ are given in table \ref{Pontryagin table}. We can see that if we specify the ALF result to Taub-NUT we recover the classical results of \cite{Grimm:2018weo} as expected.


\subsection{Levels of non-abelian flavour symmetries}

In order to compute the levels of the non-abelian flavour symmetries we should perform similar manipulations as in the previous section. We reduce the action on the compact part of the black string solution including the vectors corresponding to the non-abelian flavour symmetries, and extract the coefficients of the Chern--Simons terms of the gauge fields. Since these fields have trivial profile in the black string solution we take the trivial reduction ansatz $\hat A^i=A^i$ for the gauge-fields, where now the $A^i$ just depend on $M_3$. The only potential source for Chern--Simons terms for the gauge fields are the terms
\begin{equation}
\frac{1}{2(2\pi)^3} \sum_{i} \eta_{\alpha\beta} s_{i}^{\alpha}\int_{M_6} \hat{B}^{\beta}_{\Gamma}\wedge  \tr \hat{F}^{i}\wedge \hat{F}^{i}
\end{equation}
in the 6d action \eqref{action}. Reducing these terms with the ansatz for the gauge fields results in
\begin{align}
\frac{1}{2(2\pi)^3} \sum_{i} \eta_{\alpha\beta} s_{i}^{\alpha}\int_{M_6} \hat{B}^{\beta}_{\Gamma}\wedge  \tr \hat{F}^{i}\wedge \hat{F}^{i}
&=-\frac{1}{2(2\pi)^3}\sum_{i} \eta_{\alpha\beta}s_{i}^{\alpha}\int_{M_6}\hat{G}_{\Gamma}^{\beta} \wedge \omega^{\text{CS}}(\hat A^{i})\nonumber\\
&= \frac{1}{2(2\pi)^3}\sum_{i} \eta_{\alpha\beta} s_{i}^{\alpha} \int_{M_3}\omega^{\text{CS}}(A^{i})\int_{M^{\text{sph}}} \frac{1}{|\Gamma|}\hat{G}^{\beta} \nn\\
&=-\frac{1}{4\pi} \sum_{i}\eta_{\alpha\beta}s_{i}^{\alpha} Q^{\beta} \int_{M_3} \omega^{\text{CS}}(A^{i})\, . 
\end{align}
For such a left-moving\footnote{In our conventions a positive definite coefficient for a non-abelian Chern--Simons term implies that the current is right-moving, similarly a negative definite coefficient implies that it is left-moving.} non-abelian current algebra the level is defined via 
\be
S_{\text{CS}}\supset -\frac{1}{4\pi} \sum_{i} k_{G_{i}}^{\text{class}}\int_{M_3}\omega^{\text{CS}}(A^{i})\,.
\ee
Therefore, the level associated to the gauge group factor $G_{i}$ is 
\be
k_{G_{i}}^{\text{class}}=\eta_{\alpha\beta}s_{i}^{\alpha} Q^{\beta}\equiv \Big( C-\frac{1}{4} p_{1}(M_{\Gamma}) c_{1}(B)\Big) \cdot S_{i}\, .
\ee
For the final equality we have used the relation between the microscopic and macroscopic charges in \eqref{Charge shift}.


\section{Quantum contributions to levels}\label{sec:quantumcontr}

Having determined the contributions arising from the classical action we now turn our attention to computing the quantum corrections to the levels. Such corrections arise from one-loop Chern--Simons terms induced by integrating out massive Kaluza--Klein (KK) states. We will only be interested in corrections up to but not including terms $\mathcal{O}(1)$ in the charges. The KK modes contributing to the one-loop Chern--Simons terms come from the six-dimensional fields which can lead to anomalies in the six-dimensional theory, namely the chiral fields. For the theory at hand these 6d fields are the gravitino, the spin-$\tfrac{1}{2}$ fermions in the tensor-, vector- and hypermultiplets and the (anti-)self-dual two-forms. Upon reducing to three dimensions these fields give rise to massive spin-$\tfrac{3}{2}$ and spin-$\tfrac{1}{2}$ fermions, and to massive chiral vector fields. 

We argued in section \ref{sec:classcontr} that in order to obtain the correct anomaly coefficients from supergravity, one has to perform the dimensional reduction of the six-dimensional effective action to three dimensions on the spherical part of the asymptotic geometry which surrounds the string in six dimensions. The two classes of transverse spaces which we consider in this paper, namely ALE and ALF spaces, behave qualitatively different at asymptotic infinity. For an ALE space the metric approaches a quotient of the Euclidean metric, whereas for an ALF space it approaches the metric on ${(\mathbb{R}^3 \times \t{S}^1)/\Gamma}$. In both cases the spherical geometry in the asymptotic region is topologically S$^3/\Gamma$, however whilst one is left with a finite radius circle in the asymptotic geometry of an ALF space (the Hopf fiber of the S$^3$), the S$^3/\Gamma$ in the asymptotic ALE geometry becomes infinitely large. The KK modes, whose masses are inversely proportional to the radius, therefore become massless for an ALE geometry and consequently they do not contribute to the one-loop Chern--Simons terms. In contrast, because of the finite Hopf fiber for ALF transverse spaces the KK modes remain finite and must be integrated out at low energies. It follows that one only needs to compute the one-loop corrections to the Chern--Simons levels for transverse ALF spaces and not for ALE spaces. 

In the following we assume that the squashing of the three-sphere does not alter the representation content of the KK modes under the isometry group and moreover that it does not change the signs of the KK masses of the states in the spectrum. This will effectively mean that we can compute the one-loop corrections on the round S$^3/\Gamma$. In \cite{Couzens:2019wls,Grimm:2018weo} for similar settings, this assumption was shown to lead to results which agree with the microscopic predictions.

\subsection{Kaluza--Klein spectrum}\label{sec:spectrum}

The spectrum of $\mathcal{N}=(2,0)$ supergravity on AdS$_3\times \mathrm{S}^3$ was computed in \cite{Deger:1998nm,deBoer:1998kjm} and can be truncated to the spectrum of $\cN=(1,0)$ supergravity coupled to tensor multiplets. Each mode in the spectrum admits a representation under the isometry group ${\mathfrak{so}(4)=\mathfrak{su}(2)_L\oplus \mathfrak{su}(2)_{R}}$ and has a mass of which only the sign will be important in this paper. In addition to tensor multiplets we have hypermultiplets and vector multiplets in the $\mathcal{N}=(1,0)$ 6d supergravity theory which are absent in the truncation from $\mathcal{N}=(2,0)$ supergravity. Since only the massive KK modes of the spin-$\frac{1}{2}$ fermions in these multiplets are relevant for the computation of the one-loop Chern--Simons terms in 3d we focus on these. The computation of the KK spectrum of these spin-$\frac{1}{2}$ fields requires a harmonic expansion of the linearized Dirac equation, analogous to the one for the spin-$\frac{1}{2}$ fermions in the tensor multiplets. It is therefore conceivable that the massive KK spin-$\frac{1}{2}$ fermions from the hyper- and vector multiplets fall into the same $\mathfrak{su}(2)_L \oplus \mathfrak{su}(2)_R$ representations as the KK fermions from the tensor multiplets. The only potential difference is in the sign of the KK mass which is correlated with the chirality of the parent 6d spin-$\frac{1}{2}$ fermions. The 6d fermions are given by two Weyl fermions subject to a symplectic-Majorana condition and the tensors obey a reality condition. It is computationally simpler to not impose these conditions from the outset but to impose them at a later point.

Taking these considerations into account we can infer the KK spectrum of the ${\mathcal{N}=(1,0)}$ 6d supergravity theory coupled to $n_T$ tensor multiplets, $n_V$ vector multiplets and $n_H$ hypermultiplets on AdS$_3\times \mathrm{S}^3$. It is given below in terms of representations of $\mathfrak{su}(2)_{L}\oplus \mathfrak{su}(2)_{R}\oplus \mathfrak{g}$, as $(j_{L},j_{R},\RR)^{\text{sgn}(M)}$ with the superscript denoting the sign of the mass of the field.
\begin{itemize}
\item Spin-$\tfrac{3}{2}$:
\begin{equation*}
2 \bigoplus_{j_{L}=\tfrac{1}{2}}^{\infty} \big(j_{L},j_{L}\pm \tfrac{1}{2},\boldsymbol{1}\big)^{\mp}\, .
\end{equation*}
\item
Spin-$\tfrac{1}{2}$:
\begin{align*}
&2 \bigoplus_{j_{L}=\tfrac{3}{2}}^{\infty}\big(j_{L},j_{L}\pm \tfrac{3}{2},\boldsymbol{1}\big)^{\mp}   \oplus    2 \bigoplus_{j_{L}=0}^{1}\big(j_{L}, j_{L}+\tfrac{3}{2},\boldsymbol{1}\big)^{-}	\oplus	2 \bigoplus_{j_{L}=1}^{\infty}\big(j_{L},j_{L}\pm \tfrac{1}{2},\boldsymbol{1}\big)^{\pm} \\%
\oplus &2 \big(\tfrac{1}{2},1, \boldsymbol{1}\big)^{+}\oplus 2 \big(0,\tfrac{1}{2},\boldsymbol{1}\big)^{+}\oplus 2 n_{T}\bigg[ \bigoplus_{j_L=\tfrac{1}{2}}^{\infty}\big(j_{L}, j_{L}\pm \tfrac{1}{2},\boldsymbol{1}\big)^{\pm}\oplus \big(0,\tfrac{1}{2}, \boldsymbol{1}\big)^{+}\bigg]  \\%
\oplus &2\bigoplus_{\RR} x_{\RR} \bigg[\bigoplus^{\infty}_{j_L=\tfrac{1}{2}}\big(j_{L}, j_{L}\pm \tfrac{1}{2}, \RR\big)^{\pm}\oplus \big(0,\tfrac{1}{2},\RR\big)^{+}\bigg]\\%
\oplus &2\bigoplus_{i}\bigg[\bigoplus_{j_{L}=\tfrac{1}{2}}^{\infty} \big( j_{L}, j_{L}\pm \tfrac{1}{2}, \boldsymbol{\mathrm{ad}}_{i}\big)^{\mp}\oplus \big(0,\tfrac{1}{2},\boldsymbol{\mathrm{ad}}_{i}\big)^{-}\bigg]\,.
\end{align*}
\item 
Chiral vectors:
\begin{align*}
&\bigoplus_{j_{L}=1}^{\infty} \big(j_{L}, j_{L}\pm 1, \boldsymbol{1}\big)^{\mp}\oplus \big(\tfrac{1}{2}, \tfrac{3}{2}, \boldsymbol{1}\big)^{-}\oplus \big(0,1,\boldsymbol{1}\big)^{-}\\%
\oplus n_{T}\bigg[ &\bigoplus_{j_{L}=1}^{\infty} \big(j_{L}, j_{L}\pm1, \boldsymbol{1}\big)^{\pm} \oplus \big(\tfrac{1}{2}, \tfrac{3}{2}, \boldsymbol{1}\big)^{+}\oplus \big(0,1,\boldsymbol{1}\big)^{+}\bigg]\,.
\end{align*}

\end{itemize}
For simplicity we have used the shorthand notation 
\be
\big(j_{L}, j_{L}\pm n\big)^{\pm}\equiv \big(j_{L},j_{L}+n\big)^{+}\oplus \big(j_{L},j_{L}-n\big)^{-}\, .
\ee
The hypermultiplets transform in representations $\RR$ of $\mathfrak{g}$ which come with multiplicities $x_{\RR}$. Moreover we have denoted $\boldsymbol{\mathrm{ad}}_{i}$ for the representation under $\mathfrak{g}$ where the field is in the adjoint of the $i$'th gauge group factor $G_i$ and singlet of the other group factors. Therefore,
\be\label{defnhnv}
\sum_{\RR} x_{\RR}\mathrm{dim}(\RR)=n_H\,, \qquad
\sum_{i} \mathrm{dim}(\boldsymbol{\mathrm{ad}}_{i})=n_V\,.
\ee 
Denoting the eigenvalues of the Cartans of $\mathfrak{su}(2)_{L,R}$ by $j_{L,R}^{\scriptscriptstyle{(3)}}$, the 6d reality and symplectic-Majorana conditions of the tensors and fermions respectively map modes with eigenvalues $(j_L^{\scriptscriptstyle{(3)}}, j_R^{\scriptscriptstyle{(3)}}) \to (-j_L^{\scriptscriptstyle{(3)}}, -j_R^{\scriptscriptstyle{(3)}})$. We are thus left with either the modes that satisfy $j_{L}^{\scriptscriptstyle{(3)}}\geq 0$ or the modes that satisfy $j_{R}^{\scriptscriptstyle{(3)}}\geq 0$.

Ultimately we are interested in the one-loop corrections induced by KK modes on S$^3/\Gamma$. We must therefore extract out the $\Gamma$-invariant subsector of the above spectrum. Since we need to compute the corrections for the ALF spaces only, we are interested in the projection conditions for $\Gamma$ either $\mathbb{Z}_m$ or $\mathbb{D}^*_m$. The action of the subgroup $\Gamma \subset SU(2)_L$ may be understood by elementary group theory techniques. From the action of $\Gamma$ on the fundamental representation of $SU(2)_L$ one can straightforwardly deduce its action on a general representation using that the latter can be written as a totally symmetric tensor product of the fundamental representation. One finds for $\Gamma= \mathbb{Z}_m$ that only states with $j_L^{\scriptscriptstyle{(3)}}=\frac{1}{2}m k$ for some $k \in \mathbb{Z}$ are invariant under $\mathbb{Z}_m$. The symplectic-Majorana and reality conditions refine this to $k\in \mathbb{Z}_{\geq0}$. For $\Gamma = \mathbb{D}^*_m$ we first note that $\mathbb{D}^*_{m}$ has a $\mathbb{Z}_{2m}$ subgroup. This implies, that the invariant states need to satisfy $j_L^{\scriptscriptstyle{(3)}}= m k$ for some $k \in \mathbb{Z}$. In addition to this subgroup, $\mathbb{D}^*_{m}$ has another generator that maps a state with eigenvalue  $j_L^{\scriptscriptstyle{(3)}}$ to a state with eigenvalue $-j_L^{\scriptscriptstyle{(3)}}$. By taking appropriate linear combinations of these two states we obtain a state which is invariant under the full group $\mathbb{D}^*_m$. We can therefore restrict to states that have $j_L^{\scriptscriptstyle{(3)}}=m k$ for some $k \in \mathbb{Z}_{\geq 0}$. Furthermore the symplectic-Majorana and reality conditions imply that we should only take states with $j_R^{\scriptscriptstyle{(3)}}\geq 0$. A more detailed discussion of the projection conditions can be found in appendix A of \cite{Couzens:2019wls}.


\subsection{One-loop contributions of massive KK modes to Chern--Simons terms}\label{single field loops}

The contributions of the KK modes to the $\mathfrak{u}(1)_{L}$ , $\mathfrak{su}(2)_{R}$ and gravitational Chern--Simons terms were computed in \cite{Grimm:2018weo} in the absence of vector multiplets and charged matter for Taub--Nut transverse space. Here we extend those results to include vector multiplets and charged matter and to the D-series. Since the only matter that can contribute to the coefficients of the Chern--Simons terms and that is also charged under the gauge group $G$ are spin-$\tfrac{1}{2}$ fermions, we solely need to modify the contributions from these terms. Moreover we must compute the contribution of these fields to the levels of the flavour current algebra. 

Let $j_{L}^{\scriptscriptstyle{(3)}}$ and $j_{R}$ be the quantum numbers of the Casimirs of a representation of $\mathfrak{u}(1)_{L}\oplus\mathfrak{su}(2)_{R}$. If one computes a two-point function of the currents in the theory, one sees that in integrating out the massive fields one-loop Chern--Simons terms of the form
\be
\alpha_L \int_{M_{3}} A_{L}\wedge F_{L} +\alpha_R\int_{ M_3}\omega^{\mathrm{CS}}(A_{R})+\alpha_{\text{grav}} \int_{ M_3} \omega^{\mathrm{CS}}_{\text{grav}}+\alpha_{G_i}\int_{M_3}\omega^{\text{CS}}(A^{i})\, 
\ee
are induced. The contribution of each massive field can be evaluated by either explicitly computing the one-loop Feynman diagram or by using the index theorem and appealing to anomaly inflow. For example the parity anomaly of a spin-$\tfrac{1}{2}$ field is canceled by the counter-term \cite{AlvarezGaume:1984nf}
\be
\pi \mathrm{sgn}(M) \int_{M_{3}}Q_{\tfrac{1}{2}}\big( \{ A^{i}\},\omega\big)\, ,
\ee
where 
\begin{align}
\mathrm{d}Q_{\frac{1}{2}}(\{A^{i}\}, \omega)&=\hat A(M_3) \wedge \mathrm{ch}(F)\big|_{\text{4-form}}\nonumber\\%
&=\hat A(M_3) \wedge \mathrm{ch}(F_{L})\wedge \mathrm{ch}(F_{R})\wedge \bigwedge_i  \mathrm{ch}(F^i)\big|_{\text{4-form}}\,,\nn \\
\hat A(M_3)&=1+\frac{1}{(4\pi)^2}\frac{1}{12} \mathrm{tr }\, \mathcal R \wedge \mathcal R + \dots\,,  \\
\mathrm{ch}(F^i)&=\mathrm{dim}\, \boldsymbol{\mathrm{R}}_i+ \frac{i}{2\pi} \mathrm{tr}_{\boldsymbol{\mathrm{R}}_i}\, F^i-\frac{1}{2}\frac{1}{(2\pi)^2}\mathrm{tr}_{\boldsymbol{\mathrm{R}}_i} \, F^i \wedge F^i + \dots\, .\nn
\end{align}
Let $r$ be the dimension of the $SU(2)_{R}$ representation. Since we have assumed that the group factors $G_{i}$ are simple it follows that $\tr F^{i}=0$ and therefore
\begin{align}
\mathrm{d}Q_{\frac{1}{2}}(\{A^{i}\}, \omega)=&\, \frac{r\, \mathrm{dim}\, \boldsymbol{\mathrm{R}}}{(4\pi)^2}\frac{1}{12} \mathrm{tr }\, \mathcal R \wedge \mathcal R-\frac{r}{2(2\pi)^2}\sum_i d_i (\boldsymbol{\mathrm{R}}) \mathrm{tr}_{\boldsymbol{\mathrm{R}}_i} \, F^i \wedge F^i\, \nonumber\\%
&\, +\frac{r\, \mathrm{dim}\, \RR}{4(2\pi)^2} F_{L}\wedge F_{L}-\frac{\mathrm{dim}\, \RR}{2(2\pi)^2} \tr F_{R}\wedge F_{R}\,  \nonumber\\%
=&\, \frac{r \,\mathrm{dim}\, \boldsymbol{\mathrm{R}}}{(4\pi)^2}\frac{1}{12} \mathrm{tr }\, \mathcal R \wedge \mathcal R-\frac{r}{2(2\pi)^2}\sum_i d_i (\boldsymbol{\mathrm{R}})A_{\boldsymbol{\mathrm{R}}_i }\lambda_i\, \mathrm{tr} \, F^i \wedge F^i\, \\
&\, +\frac{r \,\mathrm{dim}\, \RR}{4(2\pi)^2} F_{L}\wedge F_{L}-\frac{\mathrm{dim}\, \RR}{2(2\pi)^2} \tr F_{R}\wedge F_{R}\, ,\label{dQ}\nn
\end{align}
with 
\be
\mathrm{dim}\, {\boldsymbol{\mathrm{R}}}=\prod_i \mathrm{dim}\, {\boldsymbol{\mathrm{R}}}_i\, ,\qquad d_i (\boldsymbol{\mathrm{R}})=\prod_{j \neq i} \mathrm{dim}\, {\boldsymbol{\mathrm{R}}_j}\, ,
\ee
and the group theoretic objects, $\lambda_{i}$ and $A_{\RR_i}$ are as introduced in section \ref{6d sugra}. We used that the generator of $\mathfrak{u}(1)_L$ is given in terms of the Pauli matrices by $-\tfrac{i}{2}\sigma_3$. It follows that the contribution of a single spin-$\tfrac{1}{2}$ fermion  in the representation $\RR$ of $G$ gives a contribution of 
\be
\alpha^{(1/2)}_{G_{i}}=-\frac{r}{8\pi} \mathrm{sgn}(M) d_{i}(\RR) A_{\RR_{i}} \lambda_{i}
\ee
to the Chern--Simons term corresponding to the non-abelian flavour symmetry $G_{i}$. The contribution to the universal sector is\footnote{In translating between the trace in the Chern--Simons terms used here and the one used in the section on the classical contributions a field in the representation $(j^{\scriptscriptstyle{(3)}}_{L},j_{R})$ picks up a factor of $2 (j_{L}^{\scriptscriptstyle{(3)}})^2$ for left-moving Chern--Simons terms and a factor of $\tfrac{2}{3}j_{R}(j_{R}+1)(2 j_{R}+1)$ for right-moving Chern--Simons terms. This is the analogue of $\lambda_{i}$ appearing earlier. Moreover we substitute the $r=2 j_{R}+1$.}
\begin{align}
\alpha^{(1/2)}_{L}&=\frac{1}{8 \pi}\mathrm{sgn}(M) \mathrm{dim}\RR \, (2 j_{R}+1)\big(j_{L}^{(3)}\big)^2\, , \nn\\
\alpha_{R}^{(1/2)}&=-\frac{1}{12 \pi}\mathrm{sgn}(M) \mathrm{dim}\RR \, j_{R}(j_{R}+1)(2 j_{R}+1)\, ,\\
\alpha^{(1/2)}_{\mathrm{grav}}&= \frac{1}{192 \pi} \mathrm{sgn}(M) \mathrm{dim}\RR \, (2 j_{R}+1)\, .\nn
\end{align}
Finally, since the other fields are singlets of the gauge group $G$ their contributions are precisely given by the ones in \cite{Grimm:2018weo}. We collect the contributions of the single fields in table \ref{one-loop contributions}.

\renewcommand{\arraystretch}{1.8}
\begin{table}[h]
\begin{center}
\begin{tabular}{C{0.9cm}| C{4.7cm} C{3.7cm}  C{3.7cm}}
\specialrule{.07em}{0.05em}{-.12em}

				& spin-$\tfrac{1}{2}$ 		& 			spin-$\tfrac{3}{2}$		&			chiral vectors\\
\hline
$\alpha_{L}$      & $\frac{1}{8 \pi}\mathrm{sgn}(M) \mathrm{dim}\,\RR\, r \big(j_{L}^{(3)}\big)^2$ & $\frac{3}{8\pi} \mathrm{sgn}(M) r \big(j_{L}^{(3)}\big)^2$	&$-\frac{1}{4\pi} \mathrm{sgn}(M) r \big(j_{L}^{(3)}\big)^2$\\

$\alpha_{R}$ & $-\frac{1}{12\pi} \mathrm{sgn}(M)\mathrm{dim}\,\RR\, \lambda_{R}$	&	$- \frac{1}{4\pi}\mathrm{sgn}(M)\lambda_{R}$   &   $\frac{1}{6 \pi} \mathrm{sgn}(M) \lambda_{R}$\\

$\alpha_{\mathrm{grav}}$ &$\frac{1}{192 \pi} \mathrm{sgn}(M) \mathrm{dim}\,\RR\, r$& $-\frac{7}{64 \pi} \mathrm{sgn}(M) r$ &$\frac{1}{48\pi} \mathrm{sgn}(M) r$\\

$\alpha_{G_{i}}$ &  $ -\frac{1}{8\pi} \mathrm{sgn}(M) d_{i}(\RR) A_{\RR_{i}} \lambda_{i}\, r$&$0$&$0$\\
\specialrule{.07em}{0.05em}{0em}
\end{tabular}
\end{center}
\caption{One-loop contributions of a single field in the representation $(j_{L}^{\scriptscriptstyle{(3)}}, j_{R}, \RR)$ of $\mathfrak{u}(1)_{L}\oplus\mathfrak{su}(2)_{R}\oplus \mathfrak{g}$ to the Chern--Simons terms. We have used the shorthand $\lambda_{R}= j_{R}(j_{R}+1) (2 j_{R}+1)$ and the dimension of the $SU(2)_{R}$ representation $r=2j_{R}+1$.}
\label{one-loop contributions}
\end{table}

In order to compute the one-loop corrections to the levels $k_L$ of $U(1)_L$, $k_R$ of $SU(2)_R$, $k_{G_i}$ of $G_i$ and $c_L-c_R$ multiplying the gravitational Chern--Simons term in 3d, we have to sum the single field contributions in table \ref{one-loop contributions} over the KK towers given in section \ref{sec:spectrum}. In performing the sums over the infinite towers of KK states one encounters sums of the form
\begin{equation}
\sum_{n=1}^\infty n^k\, ,
\end{equation}
which are divergent and need regularizing. We employ Zeta-function regularization to obtain finite results for the one-loop Chern--Simons terms. It was shown in \cite{Corvilain:2017luj,Corvilain:2020tfb}, that Zeta-function regularization gives the correct result for the constant, field independent one-loop corrections to Chern--Simons terms, while extra care has to be taken concerning the field-dependent part. Since we start from an anomaly free 6d F-theory model, Zeta-function regularization provides a safe shortcut in the computation of the three-dimensional Chern--Simons coefficients we are interested in. The details of the computation of the sums over the Kaluza--Klein towers are spelled out in appendix \ref{loop sums}. In these computations we use the identities \eqref{grav anomaly}, \eqref{nT} and \eqref{c.s} to express the final results entirely in terms of $m$, $c_1(B)$ and $S_i$. In the following we will give the results for the two relevant cases $\Gamma=\mathbb{Z}_m$ and $\Gamma=\mathbb{D}^*_m$, i.e. we present the one-loop corrections induced by massive KK modes on S$^3/\Gamma$.

\paragraph{A-series (${\Gamma=\mathbb{Z}_m}$).} We first turn our attention towards the case where $\Gamma=\mathbb{Z}_m$, which corresponds to a string propagating in 6d probing a transverse Taub-NUT space with NUT charge $m$. We are interested in contributions to $k_{L, R}$, $k_{G_i}$ and $c_L-c_R$, which scale with the charge $m$ and we will neglect all $\cO(1)$ contributions which do not depend on the charges. As described in section \ref{sec:spectrum}, we need to sum over states which satisfy the projection condition $j_L^{\scriptscriptstyle{(3)}}=\frac{1}{2}m k$ for $k \in \mathbb{Z}_{\geq 0}$. The explicit sums over all states can be found in appendix \ref{loop sums} and are performed by first summing over all representations, which contain a state with $j_L^{\scriptscriptstyle{(3)}}=\frac{1}{2}m k$ for generic $k$. That is, we sum over representations with $j_L = \frac{1}{2} m k,$ $\frac{1}{2}m k +1, \dots$, and then sum over all values of $k$. After applying the identities \eqref{grav anomaly}, \eqref{nT} and \eqref{c.s} we obtain the corrections
\begin{align}
 k_L^{\text{loop}}&=-\frac{m^3}{8}c_1(B)\cdot c_1(B)\, ,\nn\\
 k_R^{\text{loop}}&=\frac{m^3}{24} c_1(B)\cdot c_1(B)+\frac{m}{3}c_1(B)\cdot c_1(B)+m\, ,\nn\\
  k_{G_{i}}^{\mathrm{loop}}&= \frac{m}{2} c_{1}(B)\cdot S_{i}\,,\label{loop results A} \\
 (c_L- c_R)^{\text{loop}}&=2m c_1(B) \cdot c_1(B)+6m\, , \nn
\end{align}
to the levels up to $\cO(1)$ corrections which are independent of $m$.

\paragraph{D-series (${\Gamma=\mathbb{D}^*_m}$).}  For this case we do not have a left level to compute. In section \ref{sec:spectrum} we found that we need to sum over states with $j^{\scriptscriptstyle{(3)}}_L=m k$ for $k \in \mathbb{Z}_{\geq 0}$ and $j^{\scriptscriptstyle{(3)}}_R \geq 0$. It is clear that the first restriction can be imposed by simply replacing $m \to 2m$ in the one-loop results for the A-series \eqref{loop results A}. The second condition may be effectively implemented by dividing the result of that replacement by a factor two. This simple procedure gives the sums for the case $\Gamma=\mathbb{D}^*_m$ up to terms which are independent of $m$. Since this is sufficient for our purposes, we immediately obtain the result for $\Gamma=\mathbb{D}^*_m$: 
\begin{align}
 k_R^{\text{loop}}&=\frac{m^3}{6} c_1(B)\cdot c_1(B)+\frac{m}{3}c_1(B)\cdot c_1(B)+m\, ,\nn\\
 k_{G_{i}}^{\text{loop}}&= \frac{m}{2} c_{1}(B)\cdot S_{i}\,,\label{loop results D}  \\
 (c_L- c_R)^{\text{loop}}&=2m c_1(B) \cdot c_1(B)+6m\, ,\nn
\end{align}
again, up to terms independent of the charge $m$.

\section{Summary}\label{summary}

In this paper we have studied the central charges and levels corresponding to black strings in F-theory compactified on an elliptically fibered Calabi--Yau threefold CY$_3$. The strings arise from wrapping D3-branes, living in the asymptotic geometry $\mathbb{R}\times M_\Gamma \times \mathrm{S}^1\times \mathrm{CY}_3$, on the curve $C$ in the base of the CY$_3$ and on the S$^1$. The space $M_\Gamma$ transverse to the string is either taken to be asymptotically locally Euclidean (ALE) or asymptotically locally flat (ALF). These spaces are characterized by the choice of a freely acting discrete subgroup $\Gamma \subset SU(2)$. Living on the strings are 2d $\mathcal{N}=(0,4)$ SCFTs which result from compactifying the woldvolume theory of the branes on $C$. These SCFTs have associated central charges and levels which we have computed from the macroscopic 6d $\mathcal{N}=(1,0)$ supergravity theory which is the low energy limit of F-theory on CY$_3$.

The isometry group of the ALE and ALF spaces is $U(1)_L \times SU(2)_R$ for the A-series and $SU(2)_R$ for the D- and E-series in the ADE classification of subgroups of $SU(2)$. This isometry group corresponds to a current algebra in the dual SCFT with associated levels $k_{L,R}$. Degenerations in the fiber of the CY$_3$ leading to singularities of the total space yield vector multiplets in the 6d supergravity theory which lead to gauge symmetries of the 6d bulk theory. From the perspective of the string, this leads to non-abelian flavour symmetries of the 2d SCFT each of which has an associated level.

Macroscopically the levels and central charges correspond to the coefficients of Chern--Simons terms in the effective 3d action that is obtained by reducing the 6d supergravity theory on the spherical part of the black string solutions. We have determined these Chern--Simons terms including one-loop contributions arising from integrating out massive Kaluza--Klein modes. We have performed the reduction at asymptotic infinity in order to take into account the contributions of degrees of freedom living outside of the horizon. In addition, a necessary shift in the identification of macroscopic and microscopic charges arising from the Green--Schwarz--Sagnotti--West term in the pseudo-action has been included.

When the transverse space is ALE the central charges and levels are given by the coefficients of the Chern--Simons terms resulting from the reduction of the 6d classical action on the spherical part of the black string geometry. We have computed this in section \ref{sec:classcontr} and the result is
\begin{align}\label{summale}
k_{L}&= \frac{1}{2}|\Gamma|\Big( C -\frac{1}{4} p_{1}(M_{\Gamma}) c_{1}(B)\Big)^2-\frac{1}{2} c_{1}(B)\cdot\Big( C -\frac{1}{4} p_{1}(M_{\Gamma}) c_{1}(B)\Big)\, ,\nonumber \\
k_{R}&= \frac{1}{2}|\Gamma|\Big( C -\frac{1}{4} p_{1}(M_{\Gamma}) c_{1}(B)\Big)^2+\frac{1}{2} c_{1}(B)\cdot\Big( C -\frac{1}{4} p_{1}(M_{\Gamma}) c_{1}(B)\Big)\, ,\nn\\
k_{G_{i}}&= \Big( C-\frac{1}{4} p_{1}(M_{\Gamma}) c_{1}(B)\Big) \cdot S_{i}\, ,\\
c_{L}-c_{R}&= 6 c_{1}(B)\cdot \Big( C -\frac{1}{4} p_{1}(M_{\Gamma}) c_{1}(B)\Big)\, ,\nn 
\end{align}
where $|\Gamma|$ and $p_1(M_\Gamma)$ can be found in table \ref{Pontryagin table}. The left level is only relevant for the A-series. As discussed in section \ref{sec:classcontr}, the left- and right-moving central charges follow from \eqref{summale} by using the relation $c_R=6k_R$ which is valid when one can identify $SU(2)_R$ with the right-moving R-symmetry.

When the transverse space is ALF one also has to include the one-loop contributions derived in section \ref{sec:quantumcontr}. For the A-series, i.e. $M_\Gamma=\mathrm{TN}_m$, we find that the final result is
\begin{align}
k_{L}&=  \frac{1}{2}m C^2-\frac{1}{2}m^2  c_{1}(B)\cdot C \, ,\nonumber \\
k_{R}&= \frac{1}{2}m C^2 -\frac{1}{2}m^2  c_1(B) \cdot C+\frac{1}{6}m^3 c_1(B)^2 + c_{1}(B)\cdot C -\frac{1}{6}m c_{1}(B)^2+m\, ,\nn\\\
k_{G_{i}}&=  C \cdot S_{i}\, ,\\
c_{L}-c_{R}&= 6 c_{1}(B)\cdot C- m  c_{1}(B)^2+6m\, .\nn
\end{align}
Alternatively when the transverse space $M_\Gamma$ is ALF with $\Gamma$ given by the D-series, the levels and central charges are\footnote{These results are the sums of the classical and quantum contributions for the D-series up to terms of $\mathcal{O}(1)$ in the charges.}
\begin{align}\label{finaldseries}
k_{R} & =  
2m C^2 -( 2m^2+6m-1)c_1(B)\cdot C +\frac{1}{6}(4m^3+18m^2+26m)c_1(B)^2+m\,, \nonumber\\
k_{G_{i}}&= C \cdot S_{i}\, ,\\
c_{L}-c_R & =  6 c_1(B)\cdot C- m c_1(B)^2+6m\,.\nn
\end{align}
All results in this section are up to terms of $\mathcal{O}(1)$ in the charges.


\section{Discussion}\label{sec:discu}

The results in this paper have been obtained from purely macroscopic computations, it would be very interesting to reproduce the central charges and levels from a microscopic computation. This is possible for the central charges and levels corresponding to the isometry group of $M_\Gamma$ when the transverse space is either Taub-NUT \cite{Bena:2006qm} or $\mathbb{R}^4$ \cite{Haghighat:2015ega,Lawrie:2016axq}. For Taub-NUT one can use the dual M-theory setting while for $\mathbb{R}^4$ one can work directly with the woldvolume theory of the D3-brane. In the latter case one compactifies the theory living on a D3-brane on the curve $C$ in the presence of a varying axio-dilaton which requires one to perform a topological duality twist \cite{Martucci:2014ema} which was further studied in \cite{Assel:2016wcr,Lawrie:2016axq}. The central charges and levels are computed in terms of anomaly coefficients or by spectrum counting in the resulting 2d theory.\footnote{The microscopic computation of the central charges/anomaly coefficients misses subleading contributions from D3-D7 modes. These contributions may again be obtained from the dual M5-brane picture \cite{Haghighat:2015ega,Lawrie:2016axq}.} Generalizing the latter computation to arbitrary ALE $M_\Gamma$ is not straightforward. The worldvolume theory of a D3-brane probing $M_\Gamma$ is known \cite{Douglas:1996sw,Johnson:1996py}, and its construction relies on performing a quotient of the non-abelian theory living on a stack of D3-branes. However the topological duality twist is not yet understood well enough for non-abelian gauge theories without appealing to M-theory and therefore at present has no obvious application to these theories. It is interesting to note that the difficulties for ALE transverse spaces are purely the result of the non-trivial elliptic fibration since if one considers type IIB settings without 7-branes one can study these strings both macroscopically and microscopically. We have pursued this for type IIB compactifications on K3 in \cite{Couzens:2019wls}. 

One also encounters problems if one wants to compute the levels for the ALF D-series microscopically.  In fact the situation is worse than for ALE spaces since the four-dimensional parent theory is not even known in this case. Instead one can try to mimic the progress made in the A-series ALF case by using an M-theory picture and performing a similar computation as in \cite{Maldacena:1997de}. The D-series theory admits a dual M-theory realization if one also introduces orientifold M5-branes. However the burden to bare by introducing orientifolds is too high with the available technology and the counting of states is currently intractable. A potential avenue for obtaining the microscopic results is via anomaly inflow in F-theory, see \cite{Lawrie:2018jut,Bah:2020jas} for the current status of this line of research. 

An alternative approach, applicable to all the settings considered here, is to dualize along the circle wrapped by the D3-brane and then uplift to M-theory. One obtains an M2-brane system wrapped on a curve in the elliptically fibered CY$_3$ and probing the ALE or ALF space. For transverse space $\mathbb{R}^4$ the spectrum counting of the M2-brane states was performed in \cite{Vafa:1997gr}, however for transverse ALE or ALF spaces this has not been considered so far. As noted in \cite{vanBeest:2020vlv} there are potential subtleties in matching the central charge of the 2d SCFT with the computation of the partition function of the dual 1d SQM theory living on the M2-branes.

We have considered a general class of Calabi--Yau threefolds admitting a single rational section, known as the zero section $\sigma_0: B \rightarrow \text{CY}_3$, whose existence distinguishes between an elliptic fibration rather than a genus-one fibration without section. The zero section maps each point of the base to the zero-point on the elliptic fiber. One may consider additional sections to the zero section which form a finitely generated abelian group called the free part of the Mordell--Weil group. Up to potential subtleties which are beyond the scope of this discussion, these sections give rise to $U(1)$ gauge factors in the six-dimensional supergravity theory in addition to the non-abelian group factors considered here. These may then again act as flavour symmetries on the strings which we considered in this paper. One could therefore extend our results to include these $U(1)$ factors. Computing the (classical and quantum) contributions to the anomaly coefficients of these $U(1)$ flavour symmetries of the strings appears straightforward with the methods presented here.

An additional direction one can follow is to consider Calabi--Yau threefolds which do not admit crepant resolutions. Implicitly the Calabi--Yau threefold used in the compactification to six-dimensional supergravity admits a crepant resolution. This implies that the resolution does not change the canonical bundle of the space and therefore the space remains Calabi--Yau. However there exist Calabi--Yau threefolds which admit non-crepant resolvable singularities. From the physics point of view the existence of non-crepant resolvable singularities implies that there is matter which is not charged under any of the massless gauge fields in the five-dimensional effective action arising from M-theory on the Calabi--Yau threefold, see e.g.~\cite{Arras:2016evy}. Instead the matter may be charged under a massive $U(1)$ or a discrete $\mathbb{Z}_{k}$ symmetry. The existence of this additional matter will affect the one-loop computations performed here.

A further interesting direction has already been alluded to in section \ref{black string}. We noted that if one considers a non-trivial profile for the vector multiplets in the string background such that they generate instanton configurations living on the ALE or ALF space, one introduces an additional shift of the macroscopic charge. From the F-theory perspective this arises from gauge instantons living on the 7-branes which induce D3-brane charge localized on the 7-brane worldvolume. This problem has not been attempted even in the simplest case of $\mathbb{R}^4$ transverse space. Suitable gauge instanton configurations on both $\mathbb{R}^4$ and Taub--NUT are known in the literature, therefore leaving the possibility to extend the settings discussed in the paper to include these instantons.

\paragraph*{Acknowledgements.}

It is a pleasure to thank Pierre Corvilain and Thomas Grimm for valuable discussions. 

This work was supported in part by the D-ITP consortium, a program of the Netherlands Organization for Scientific Research (NWO) that is funded by the Dutch Ministry of Education,
Culture and Science (OCW), and by the NWO Graduate Programme.

\newpage


\appendix

\section{6d to 3d reduction higher derivative term \label{sec: higher der calculation}}\label{appendixredhigh}

In this appendix we show how one obtains the result \eqref{resulthigher}. That is, we determine the 3d Chern--Simons terms resulting from the integral\footnote{This integral is also computed in appendix A of \cite{Grimm:2018weo}, but we include it here so that the paper is self-contained.}
\begin{equation}\label{higherrrr}
\int_{M^{\mathrm{sph}}}\hat{G}^{\alpha}\wedge\hat{\omega}_{\mathrm{grav}}^{\mathrm{CS}}\,.
\end{equation}
The integral is over the spherical part of the black string solution \eqref{blackstringsolution} with $\mathrm{d}s^2(M_\Gamma)$ replaced by the covering space in \eqref{coveringspace}. 

In order to perform this computation, we first discuss a few details of this black string solution. Its explicit form is given by 
\be\label{blackstringsolutioncov}
\dd \hat{s}_6^2= 2 H^{-1}\dd u\Big (\dd v  +\frac{1}{2} \mathcal{F}\dd u \Big) + H \dd s^2_4\, ,
\ee 
with $\dd s^2_4$ the metric in equation \eqref{coveringspace}. Here 
\begin{equation}
H=\big(\eta_{\alpha \beta}W^\alpha W^\beta \big)^{1/2}\,,
\end{equation}
and the functions $\mathcal{F}$ and $W^\alpha$ are given by
\begin{align}
\mathcal{F}&=1-\frac{n}{r}\,,\nn \\
W^\alpha&=w_\infty^\alpha+\frac{Q^\alpha}{4r}\,.
\end{align}
The constants $w_\infty^\alpha$ satisfy $\eta_{\alpha \beta}w_\infty^\alpha w_\infty^\beta=1$ in order to obtain the correct asymptotic behaviour. Furthermore, the scalars in the tensor multiplets take the form
\begin{equation}
\hat \jmath^\alpha=\frac{W^\alpha}{H}
\ee
and the three-forms are given by
\begin{equation}
\hat G^\alpha=-\dd v\wedge \dd u \wedge \dd(W^\alpha H^{-2})-\ast_4 \dd(W^\alpha)\,.
\end{equation}
Here $\ast_4$ denotes the Hodge dual with respect to the metric $\dd s_4^2$.

Having given the necessary notation and conventions we turn our attention to evaluating the integral in \eqref{higherrrr}. We begin by gauging the isometries of the base space of the black string and decompose the spin connection of the resultant metric ansatz to determine the part leading to 3d Chern--Simons terms. We denote indices of the non-spherical part $M_{3}$ of the black string solution by $\tilde{a}=1,2,3$ with corresponding vielbein $\hat{e}^{\tilde{a}}$. Additionally, we denote by $\hat{\omega}_{\tilde{a}\tilde{b}}$ the components of the
spin connection $\hat{\omega}_{M_{3}}$ with respect to
the vielbein $\hat{e}^{\tilde{a}}$ of $M_{3}$ and by $\hat{\omega}_{ab}$
the components of the spin connection $\hat{\omega}_{\mathrm{sph}}$
with respect to the vielbein $\hat{e}^{a}$ of the spherical part
of the black string solution. A vielbein of the ansatz is then given by $e^{\tilde{a}}\equiv\hat{e}^{\tilde{a}},$
$e^{a}$ (see equation \eqref{vielbeinans}) and the corresponding spin connection is \cite{Duff:1986hr}
\begin{equation}\label{spinconnectionansatz}
\omega_{\tilde{a}\tilde{b}}  =  \hat{\omega}_{\tilde{a}\tilde{b}}+\frac{1}{2}F_{\tilde{a}\tilde{b}}^{i}K_{c}^{i}e^{c}\,,\qquad
\omega_{\tilde{a}b} =  \frac{1}{2}F_{\tilde{a}\tilde{c}}^{i}K_{b}^{i}\hat{e}^{\tilde{c}}\,, \qquad
\omega_{ab} = \hat{\omega}_{ab}+\big(\hat{\nabla}_{a}K_{b}^{i}\big)A^{i}\,, 
\end{equation}
where the sum over $i$ runs over the gauge fields corresponding to the symmetries of the transverse space. From the gravitational Chern--Simons form,
\begin{equation}
\hat{\omega}_{\mathrm{grav}}^{\mathrm{CS}}= \tr \Big(\omega\wedge\mathrm{d}\omega+\frac{2}{3}\omega^{3}\Big),
\end{equation}
one sees that all 3d Chern--Simons terms can be obtained by restricting \eqref{spinconnectionansatz} to
\begin{equation}
\omega_{\tilde{a}\tilde{b}}  =  \hat{\omega}_{\tilde{a}\tilde{b}}\,,\qquad
\omega_{\tilde{a}b}  =  0\,,\qquad
\omega_{ab}  =  \hat{\omega}_{ab}+\big(\hat{\nabla}_{a}K_{b}^{i}\big)A^{i}\,. 
\end{equation}
Since this connection is a direct sum, the gravitational Chern--Simons form can be written as the sum of two Chern--Simons forms, i.e.
\begin{equation}\label{sumchernsimons}
\hat{\omega}_{\mathrm{grav}}^{\mathrm{CS}}=\omega_{\mathrm{grav}}^{\mathrm{CS}}+\omega^{\mathrm{CS}}(X)\,.
\end{equation}
Here $\omega_{\mathrm{grav}}^{\mathrm{CS}}$ is the gravitational Chern--Simons form of $M_{3}$ and 
\begin{equation}
\omega^{\mathrm{CS}}(X) =  \tr \Big(X\wedge\mathrm{d}X+\frac{2}{3}X^{3}\Big)
\end{equation}
is the Chern--Simons form corresponding to the connection $X$ with components $\hat{\omega}_{ab}+\big(\hat{\nabla}_{a}K_{b}^{i}\big)A^{i}$.

The part of $\hat{G}^{\alpha}$ in the ansatz \eqref{ansatz three-form} leading to 3d Chern--Simons terms is
\begin{equation}
-Q^{\alpha}(2\pi)^{2}|\Gamma|(e_{3}-\chi_{3})\,.
\end{equation}
The three-form $e_3$ is equal to \cite{Hansen:2006wu}
\begin{equation}
e_3=
\begin{dcases}
\frac{1}{2\pi^2}\Big[e^1\wedge e^2\wedge e^3+\frac{1}{2}K_{Ra}^I e^a\wedge F_R^I-\frac{1}{2}K_{La}e^a\wedge F_L\Big]\quad &\text{A-series}\,,\vspace{.2cm}\\[0.2cm]
\frac{1}{2\pi^2}\Big[e^1\wedge e^2\wedge e^3+\frac{1}{2}K_{Ra}^I e^a\wedge F_R^I\Big]\quad &\text{D- and E-series}\,,
\end{dcases}
\end{equation}
where the vielbein $e^i$ can be found in \eqref{vielbeinans} and one has to take the limit $r\rightarrow 0$. The three-form $\chi_{3}$ has all its legs on the non-spherical part $M_{3}$ such that when wedged with $\hat{\omega}_{\mathrm{grav}}^{\mathrm{CS}}$ we only get a contribution of $\omega^{\mathrm{CS}}\left(\hat{\omega}_{\mathrm{sph}}\right)$.
Using the expansion \eqref{sumchernsimons}, the part of the integral \eqref{higherrrr} of interest to us can be written as
\begin{equation}
\int_{M^{\mathrm{sph}}}\hat{G}^{\alpha}\wedge\hat{\omega}_{\mathrm{grav}}^{\mathrm{CS}} \supset -Q^{\alpha}(2\pi)^{2}|\Gamma| \int_{M^{\mathrm{sph}}}\Big[e_{3} \wedge\omega_{\mathrm{grav}}^{\mathrm{CS}}+e_{3}\wedge\omega^{\mathrm{CS}}(X)-\chi_{3}\wedge\omega^{\mathrm{CS}}(\hat{\omega}_{\mathrm{sph}})\Big]\,.
\label{eq:integral appendix}
\end{equation}
Evaluating these three integrals results in\footnote{The second and third integral were computed using Mathematica.}
\begin{align}
\int_{M^{\mathrm{sph}}}e_{3}\wedge\omega_{\mathrm{grav}}^{\mathrm{CS}} & =  -\omega_{\mathrm{grav}}^{\mathrm{CS}}\,,\nonumber \\
\int_{M^{\mathrm{sph}}}e_{3}\wedge\omega^{\mathrm{CS}}(X) & = 
\begin{cases} 2A_{L}\wedge F_{L} \qquad &\text{A-series} \,,\\
0 \qquad &\text{D- and E-series}\,,
\end{cases}\\
\int_{M^{\mathrm{sph}}}\chi_{3}\wedge\omega^{\mathrm{CS}}(\hat{\omega}_{\mathrm{sph}}) & =  \frac{1+4v_\infty r+10v_\infty^2r^2+8v_\infty^3r^3+2v_\infty^4 r^4}{(1+v_\infty r)^4} \times 16\pi^{2}\chi_{3}\,.\nonumber 
\end{align}
Using the definition of $\chi_3$ in equation \eqref{chi3}, we find that the integral \eqref{higherrrr} leads to the following 3d Chern--Simons terms:
\begin{align}
\int_{M^{\mathrm{sph}}}\hat{G}^{\alpha}\wedge\hat{\omega}_{\mathrm{grav}}^{\mathrm{CS}} & \supset  Q^{\alpha}\left(2\pi\right)^{2}|\Gamma|\Biggl[\omega_{\mathrm{grav}}^{\mathrm{CS}}-\frac{1+4v_\infty r+2v_\infty^2 r^2}{(1+v_\infty r)^4}A_{L}\wedge F_{L} \\
  & \qquad \qquad \qquad \quad+2\frac{1+4v_\infty r+10v_\infty^2 r^{2}+8v_\infty^3 r^{3}+2v_\infty^4 r^{4}}{(1+v_\infty r)^{4}}\omega^{\mathrm{CS}}(A_{R})\Biggl]\,.\nn
\end{align}
One has to set $A_L=0$ for the D- and E-series. This is the result as presented in \eqref{resulthigher}.


\section{One-loop corrections to 3d Chern--Simons terms}\label{loop sums}

In this appendix we compute the one-loop corrections to the Chern--Simons terms. We perform the sums of the contributions of the separate fields given in table \ref{one-loop contributions} over the spectrum given in section \ref{sec:spectrum}. We will focus on ALF spaces for the A-series, i.e.~transverse Taub-NUT spaces with NUT charge $m$. All the results we give in this appendix are up to $\cO(1)$ contributions. As explained in the main text the one-loop corrections for the case of the ALF spaces in the D-series can be obtained from the A-series results by a simple procedure. Since our focus will be on contributions to the Chern--Simons levels which scale with the charges, we only need to consider the infinite towers of states in the spectrum and can neglect the isolated representations. The relevant part of the spectrum is therefore given by
\begin{itemize}
\item Spin-$\tfrac{3}{2}$:
\begin{equation*}
2 \bigoplus_{j_{L}=\tfrac{1}{2}}^{\infty} \big(j_{L},j_{L}\pm \tfrac{1}{2},\boldsymbol{1}\big)^{\mp}\,,
\end{equation*}
\item
Spin-$\tfrac{1}{2}$:
\begin{align*}
&2 \bigoplus_{j_{L}=\tfrac{3}{2}}^{\infty}\big(j_{L},j_{L}\pm \tfrac{3}{2},\boldsymbol{1}\big)^{\mp} 	\oplus 2 \bigoplus_{j_{L}=1}^{\infty}\big(j_{L},j_{L}\pm \tfrac{1}{2},\boldsymbol{1}\big)^{\pm} \oplus 2 n_{T}\bigoplus_{j_L=\tfrac{1}{2}}^{\infty}\big(j_{L}, j_{L}\pm \tfrac{1}{2},\boldsymbol{1}\big)^{\pm}  \\%
&\oplus 2\bigoplus_{\boldsymbol{\mathrm{R}}} x_{\boldsymbol{\mathrm{R}}} \bigoplus^{\infty}_{j_L=\tfrac{1}{2}}\big(j_{L}, j_{L}\pm \tfrac{1}{2}, \RR\big)^{\pm} \oplus 2\bigoplus_{i}\bigoplus_{j_{L}=\tfrac{1}{2}}^{\infty} \big( j_{L}, j_{L}\pm \tfrac{1}{2}, \boldsymbol{\mathrm{ad}}_{i}\big)^{\mp}\,,
\end{align*}
\item 
Chiral vectors:
\begin{align*}
&\bigoplus_{j_{L}=1}^{\infty} \big(j_{L}, j_{L}\pm 1, \boldsymbol{1}\big)^{\mp}\oplus n_{T} \bigoplus_{j_{L}=1}^{\infty} \big(j_{L}, j_{L}\pm1, \boldsymbol{1}\big)^{\pm} \,.
\end{align*}
\end{itemize}

\paragraph{Level $\boldsymbol{k_R}$ of the right-moving $\boldsymbol{SU(2)_R}$ current algebra.} In the following we compute the one-loop correction to the Chern--Simons level $k_R$ up to $\cO(1)$ contributions. We compute these contributions individually for the spectra of massive spin-$\frac{3}{2}$, spin-$\frac{1}{2}$ and chiral vectors in 3d. Summing these contributions, we obtain the total one-loop correction. To project on the $\mathbb{Z}_m$-invariant states in the summations over all states, we must impose $j^{\scriptscriptstyle{(3)}}_L=\frac{1}{2}m k$ for $k \in \mathbb{Z}_{\geq 0}$.  For the computation of the one-loop Chern--Simons level $k_R$ we first use Zeta-function regularization to define
\begin{align}
\cA_R(n)&=\sum_{k=1}^\infty \sum_{j_L=\frac{1}{2}m k}^\infty \Big[(j_L+n)(j_L+n+1)(2j_L+2n+1)\nn\\
&\qquad\qquad \qquad \quad ~-(j_L-n)(j_L-n+1)(2j_L-2n+1) \Big]\label{AR}\\
&=\sum_{k=1}^\infty\Big[ k\, m \,n -\frac{1}{2}k^3 m^3 n + 2 n^3-2k \,m\, n^3\Big]=-\frac{m}{12}(n-2n^3)-\frac{m^3}{240}n-n^3\, .\nn
\end{align}
This is the infinite sum involved when summing over a tower of states of the form $\big(j_L, j_L \pm n \big)^\pm$ necessary to obtain the correction to $k_R$. For the massive spin-$\frac{3}{2}$ fermions in the spectrum and using table \ref{one-loop contributions} for the contributions of a single state the contribution to the level $k_R$ is
\begin{align}
\alpha_R^{(3/2)}&=\frac{1}{2\pi} \sum_{k=1}^\infty \sum_{j_L= \frac{1}{2} m k}^\infty \Big[\big(j_L+\tfrac{1}{2}\big)\big(j_L+\tfrac{3}{2} \big)\big(2 j_L+2 \big)-\big(j_L-\tfrac{1}{2} \big)\big(j_L+\tfrac{1}{2} \big) 2 j_L \Big]\nn\\
&=\frac{1}{2\pi} \cA_R(\tfrac{1}{2})=-\frac{1}{4\pi}\Big(\frac{m}{24}+\frac{m^3}{240} \Big)+ \cO(1)\, .\label{aR32}
\end{align}
We remind the reader that the sum over modes with $k=0$ does not contribute to the level with terms which scale with the charge $m$ and have therefore been dropped. 

Up to terms of $\mathcal{O}(1)$ we can write the contribution of the spin-$\frac{1}{2}$ fields to the Chern--Simons coefficient as
\begin{align}
\alpha_R^{(1/2)}&=\frac{1}{12\pi}\Big[ 2 \cA_R(\tfrac{3}{2})-2 \cA_R(\tfrac{1}{2})-2 n_T \cA_R(\tfrac{1}{2})\nn\\
&\qquad \qquad~ -2 \sum_{\boldsymbol{\mathrm{R}}}x_{\boldsymbol{\mathrm{R}}} \text{dim} \,{\boldsymbol{\mathrm{R}}}\, \cA_R(\tfrac{1}{2})+2 \sum_i \text{dim}\, \boldsymbol{\mathrm{ad}}_{i} \, \cA_R(\tfrac{1}{2})\Big]\nn\\
&=\frac{1}{12\pi}\Big[2 \cA_R(\tfrac{3}{2})-2 \big(1+n_T+n_H-n_V \big)\cA_R(\tfrac{1}{2})\Big]\nn\\
&=\frac{1}{4\pi} \bigg[\Big(\frac{7m}{24} -\frac{m^3}{240}\Big)+(1+n_T+ n_H-n_V)\Big(\frac{m}{72}+\frac{m^3}{720}\Big)\bigg]+\cO(1)\,,\label{aR12}
\end{align}
where we used \eqref{defnhnv}.
Finally, the contribution of the massive vectors is given by
\begin{equation}
\alpha_R^{(\text{vect})}=\frac{1}{6\pi}(n_T-1)\cA_R(1)=\frac{1}{4\pi}(n_T-1)\Big(\frac{m}{18}-\frac{m^3}{360} \Big)+\cO(1)\, .\label{aRvec}
\end{equation}
Summing \eqref{aR32}, \eqref{aR12} and \eqref{aRvec} we obtain the correction to $k_R$ which scales with $m$, namely:
\begin{align}
k_R^{\text{loop}}&= 4\pi \Big[ \alpha^{(3/2)}_R+ \alpha^{(1/2)}_R+ \alpha^{(\text{vect})}_R \Big]\nn\\
&=\frac{m^3}{720}\big(n_H-n_V-n_T-3 \big)+\frac{m}{72}\big(n_H-n_V+5 n_T+15 \big)\nn\\
&\overset{\eqref{grav anomaly}}{=} \frac{m^3}{24}(9-n_T)+\frac{m}{3} (12-n_T)\nn\\
&\overset{\eqref{nT}}{=}\frac{m^3}{24} c_1(B)\cdot c_1(B)+\frac{m}{3}c_1(B)\cdot c_1(B)+m\, .
\end{align}

\paragraph{Level $\boldsymbol{k_L}$ of the left-moving $\boldsymbol{U(1)_L}$ current algebra.} We now turn to the evaluation of the one-loop corrections to the level $k_L$ of the left-moving $U(1)_L$ current algebra. We define the sum
\begin{align}
\cA_L(n)&=\sum_{k=1}^\infty \sum_{j_L=\frac{1}{2}m k}^\infty \big(\tfrac{1}{2}m k \big)^2 \Big[ 2(j_L+n) +1-2 (j_L-n)-1 \Big]\nn\\
&=\sum_{k=1}^\infty \Big[\tfrac{1}{2}k^2 m^2 n-\tfrac{1}{2}k^3 m^3 n \Big]=-\frac{m^3}{240}n\, , 
\end{align}
which appears in the computation of the contribution of a tower with $(j_L, j_L \pm n)^\pm$ to $k_L$, as can be seen from table \ref{one-loop contributions}. For spin-$\frac{3}{2}$ fermions one finds the contribution
\begin{align}
\alpha_L^{(3/2)}=-2 \times \frac{3}{8\pi} \cA_L(\tfrac{1}{2})=\frac{1}{8\pi}\frac{m^3}{80}\, .\label{aL32}
\end{align}
Likewise one obtains for the infinite towers of massive spin-$\frac{1}{2}$ fermions the contribution
\begin{align}
\alpha^{(1/2)}_L&=\frac{1}{8\pi}\Big[ -2 \cA_L(\tfrac{3}{2})+2 \cA_L(\tfrac{1}{2})+2 n_T \cA_L(\tfrac{1}{2})\nn\\
&\qquad~~~\,+ 2 \sum_{\boldsymbol{\mathrm{R}}}x_{\boldsymbol{\mathrm{R}}}\, \text{dim}\, \boldsymbol{\mathrm{R}} \, \cA_L(\tfrac{1}{2})- 2 \sum_i \text{dim}\, \boldsymbol{\mathrm{ad}}_{i} \, \cA_L(\tfrac{1}{2})  \Big]\nn\\
&=\frac{1}{8\pi}\Big[ -2 \cA_L(\tfrac{3}{2})+2 (1+n_T +n_H-n_V) \cA_L(\tfrac{1}{2})\Big]\nn\\
&=\frac{1}{8\pi} \frac{m^3}{240}(2-n_T-n_H+n_V)\, .\label{aL12}
\end{align}
Finally, the massive vectors contribute with
\begin{align}
\alpha^{(\mathrm{vect})}_L=\frac{1}{4\pi}(1-n_T)\cA_L(1)=-\frac{1}{8\pi} \frac{m^3}{120}(1-n_T)\, .\label{aLvec}
\end{align}
Summing up the individual contributions \eqref{aL32}, \eqref{aL12} and \eqref{aLvec} we find the correction to the left level
\begin{align}
k_L^{\text{loop}}&= 8 \pi \Big[\alpha^{(3/2)}_L+\alpha^{(1/2)}_L+ \alpha^{(\text{vect})}_L \Big]=\frac{m^3}{240}(3+n_T-n_H+n_V)\nn\\
&\overset{\eqref{grav anomaly}}{=}-\frac{m^3}{8}(9-n_T)\overset{\eqref{nT}}{=}-\frac{m^3}{8}c_1(B)\cdot c_1(B)\, .
\end{align}

\paragraph{Level $\boldsymbol{c_L-c_R}$ of the gravitational Chern--Simons term.} We now determine the one-loop correction to the gravitational Chern--Simons term. We once more define the infinite sum
\begin{align}
\cA_{\text{grav}}(n)&=\sum_{k=1}^\infty \sum_{j_L=\frac{1}{2}m k}^\infty\Big[ 2(j_L+n)+1-2(j_L-n)-1\Big]\nn\\
&=\sum_{k=1}^\infty2n(1-k \, m)= \frac{m}{6}n -n\, , 
\end{align}
which appears in the sums over a tower of the form $(j_L, j_L \pm n)^\pm$ for the case of the gravitational Chern--Simons term. For the tower of massive spin-$\frac{3}{2}$ states we find
\begin{equation}
\alpha_{\text{grav}}^{(3/2)}=2 \times \frac{7}{64 \pi}\cA_{\text{grav}}(\tfrac{1}{2})=\frac{1}{96\pi}\frac{7m}{4}+\cO(1)\,.\label{c32}
\end{equation}
The infinite towers of massive spin-$\frac{1}{2}$ fermions contribute with
\begin{align}
\alpha_{\text{grav}}^{(1/2)}&=\frac{1}{96\pi}\Big[ - \cA_{\text{grav}}(\tfrac{3}{2})+\cA_{\text{grav}}(\tfrac{1}{2})+ n_T\cA_{\text{grav}}(\tfrac{1}{2})\nn\\
&\qquad ~~~~~+\sum_{\boldsymbol{\mathrm{R}}}x_{\boldsymbol{\mathrm{R}}} \, \text{dim}\,\boldsymbol{\mathrm{R}} \, \cA_{\text{grav}}(\tfrac{1}{2})-\sum_i \text{dim}\, \boldsymbol{\mathrm{ad}}_{i} \, \cA_{\text{grav}}(\tfrac{1}{2})\Big]\nn\\
&=\frac{1}{96\pi} \Big[-\cA_{\text{grav}}(\tfrac{3}{2})+(n_T+n_H-n_V+1)\cA_{\text{grav}}(\tfrac{1}{2}) \Big]\nn\\
&=\frac{1}{96\pi}\frac{m}{12}(n_T+n_H-n_V-2)+\cO(1)\, .\label{c12}
\end{align}
Lastly, the contribution of the massive vectors is given by
\begin{equation}
\alpha_{\text{grav}}^{(\text{vect})}=\frac{1}{96\pi}2 (n_T-1)\cA_{\text{grav}}(1)=\frac{1}{96\pi}(n_T-1)\frac{m}{3}+\cO(1)\, .\label{cvec}
\end{equation}
The final result is obtained by summing \eqref{c32}, \eqref{c12} and \eqref{cvec}:
\begin{align}
(c_L-c_R)^{\text{loop}}&=96\pi \Big[ \alpha_{\text{grav}}^{(3/2)}+\alpha^{(1/2)}_{\text{grav}}+\alpha^{(\text{vect})}_{\text{grav}} \Big]=\frac{m}{12} (15+5 n_T+n_H-n_V)\nn\\
&\overset{\eqref{grav anomaly}}{=}2m (12-n_T)\overset{\eqref{nT}}{=}2m c_1(B) \cdot c_1(B)+6m\, .
\end{align}

\paragraph{Levels $k_{G_i}$ of the flavour symmetries.}

The remaining one-loop correction to the levels is to the level of the non-abelian flavour symmetries. From table \ref{one-loop contributions} it is clear that the relevant infinite sum is given by 
\begin{equation}
\cA_{G_i}(n)=\cA_{\text{grav}}(n) =\frac{m}{6}n -n\, .
\end{equation}
We also note that only the massive Kaluza--Klein modes of the gauginos in the adjoint and hyperinos in the representation
\be
\RR=\bigotimes_i \RR_{i}
\ee
of the total gauge group $G=\prod_i G_i$ contribute. Summing the single field contribution over the relevant Kaluza--Klein towers we find
\begin{align}
\alpha_{G_i}&= -\frac{\lambda_i}{4\pi}\sum_{\boldsymbol{\mathrm{R}}} x_{\boldsymbol{\mathrm{R}}}\, d_i(\boldsymbol{\mathrm{R}}) \,A_{\boldsymbol{\mathrm{R}}_i} \,\cA_{G_i}(\tfrac{1}{2})
+ \frac{\lambda_i}{4\pi} A_{\boldsymbol{\mathrm{ad}}_{i}} \,\cA_{G_i}(\tfrac{1}{2}) \nonumber\\
 &=-\frac{\lambda_i}{48\pi} m \Big[\sum_{\boldsymbol{\mathrm{R}}} x_{\boldsymbol{\mathrm{R}}}\, d_i(\boldsymbol{\mathrm{R}}) \,A_{\boldsymbol{\mathrm{R}}_i}-A_{\boldsymbol{\mathrm{ad}}_{i}}\Big] + \mathcal O(1)\, .
 \end{align}
We can trade the summation over the product representation $\RR$ of $G$ for a summation over the representations of the gauge group factor $G_{i}$ by making use of the identity
\begin{equation}
 \sum_{\boldsymbol{\mathrm{R}}} x_{\boldsymbol{\mathrm{R}}}\, d_i(\boldsymbol{\mathrm{R}}) \,A_{\boldsymbol{\mathrm{R}}_i} = \sum_{\boldsymbol{\mathrm{R}}_i} x^i_{\boldsymbol{\mathrm{R}}_i} A_{\boldsymbol{\mathrm{R}}_i}\, ,
 \end{equation}
 where $x^i_{\RR_{i}}$ is the multiplicity of all hypermultiplet fermions transforming in the representation $\RR_{i}$ of $G_{i}$.  Up to $\mathcal{O}(1)$ contributions we thus find that
 \begin{equation}
\alpha_{G_i}=-\frac{\lambda_i}{48\pi} m \Big[\sum_{\boldsymbol{\mathrm{R}}_i} x^i_{\boldsymbol{\mathrm{R}}_i} A_{\boldsymbol{\mathrm{R}}_i}-A_{\boldsymbol{\mathrm{ad}}_{i}}\Big]\, .
 \end{equation}
The one-loop corrections to the levels of the non-abelian flavour symmetries are then given by
\be
 k_{G_{i}}^{\mathrm{loop}}=-4\pi \times \alpha_{G_i}= \frac{\lambda_i}{12} m \Big[\sum_{\boldsymbol{\mathrm{R}}_i} x^i_{\boldsymbol{\mathrm{R}}_i} A_{\boldsymbol{\mathrm{R}}_i}-A_{\boldsymbol{\mathrm{ad}}_{i}}\Big]\overset{\eqref{c.s}}{=} \frac{m}{2}  c_{1}(B)\cdot S_{i}\,.
\ee

\bibliographystyle{utcaps}
\bibliography{references}

\end{document}